\PassOptionsToPackage{svgnames}{xcolor}
\documentclass[graybox]{svmult}
\usepackage[utf8]{inputenc}
\usepackage{color} 
\usepackage{hyperref}
\usepackage{graphicx}
\usepackage{natbib}
\usepackage{comment}
\usepackage{mathrsfs}
\usepackage{amssymb}
\usepackage{multirow}
\usepackage{tcolorbox}
\usepackage{capt-of}
\usepackage{lipsum}
\tcbuselibrary{skins,breakable}
\hypersetup{colorlinks=true, linktoc=all, linkcolor=blue, urlcolor=blue, citecolor=blue}

\usepackage[bottom]{footmisc}
\usepackage{makeidx} 
\usepackage{mathptmx}       % selects Times Roman as basic font
\usepackage{helvet}         % selects Helvetica as sans-serif font
\usepackage{courier}        % selects Courier as typewriter font
\usepackage{type1cm}        % activate if the above 3 fonts are
\newenvironment{myblock}[1]{%
    \tcolorbox[beamer,%
    noparskip,breakable,
    colback=LightBlue,colframe=DarkBlue,%
    colbacklower=LightBlue!65!DarkBlue,%
    title=#1]}%
    {\endtcolorbox}
\newcommand{\chandra}{\textit{Chandra}}

\newcommand{\rosat}{\textit{ROSAT}}

\newcommand{\suzaku}{\textit{Suzaku}}
\newcommand{\xmm}{\textit{XMM-Newton}}

\newcommand{\swift}{\textit{Swift}}

\newcommand{\nustar}{\textit{NuSTAR}}

\newcommand{\erosita}{eROSITA}
\newcommand{\hitomi}{\textit{Hitomi}}

\title*{Modeling and Simulating X-ray Spectra}

\author{Lorenzo Ducci and Christian Malacaria}
\institute{
  Lorenzo Ducci \at Institut f\"ur Astronomie und Astrophysik T\"ubingen,
                     Kepler Center for Astro and Particle Physics, University of T\"ubingen, Sand-1, D-72076, T\"ubingen, Germany;\\
                     ISDC Data Center for Astrophysics, Universit\'e de Gen\`eve, 16 chemin d'\'Ecogia, 1290 Versoix, Switzerland;\\
                     INAF -- Osservatorio Astronomico di Brera, via Bianchi 46, 23807 Merate (LC), Italy;\\
                     \email{ducci@astro.uni-tuebingen.de}
  \and
  Christian Malacaria \at International Space Science Institute (ISSI),
                    Hallerstrasse 6, 3012 Bern, Switzerland;\\
                    \email{cmalacaria.astro@gmail.com}
                    }

\begin{document}

\maketitle

\makeatletter
\renewcommand*\l@author[2]{}
\renewcommand*\l@title[2]{}
\makeatletter

\abstract{
X-ray spectroscopy is a powerful technique for the analysis of the energy distribution of X-rays from astrophysical sources.
It allows for the study of the properties, composition, and physical processes taking place at the site of emission.
X-ray spectral analysis methods are diverse, as they often need to be tailored to the specific type of instrument used to collect the data. In addition, these methods advance together with the improvement of the technology of the telescopes and detectors.
Here, we present a compact overview of the common procedures currently employed in this field. We describe the fundamental data structure and the essential auxiliary information required for conducting spectral analysis and we explore some of the most relevant aspects related to statistical and computational challenges in X-ray spectroscopy. Furthermore, we outline some practical scenarios in the context of data reduction, modeling and fitting of spectra, and spectral simulations.}

\section*{Keywords}
Methods: data analysis; X-rays: general; Techniques: spectroscopic.

\vspace{3cm}

\let\clearpage\relax

\tableofcontents
 \markboth{\leftmark}{\rightmark}

\section{Introduction}

This chapter is intended to be an introductory and concise guide to common practices in X-ray spectral analysis. The modelling and simulation of X-ray spectra is closely linked to the characteristics of the instrument that has collected (or will collect) the data and to the specificity of the objectives that the scientist wishes to achieve with it. In addition, the methods adopted for this particular task are constantly evolving. For these reasons, this chapter is not intended to be an exhaustive collection on the subject. We refer readers who wish to delve deeper into the subject to other articles, manuals and online guides listed throughout this chapter and, for more sophisticated statistical aspects of X-ray spectral analysis, to another chapter in this volume \citep[][ and references therein]{Buchner+Boorman2022}.

In the following, we describe the data structure typical of X-ray observations and the ancillary information necessary for performing spectral analysis, and touch upon some relevant features in statistical and computational problems. We also mention some of the most useful cases for practical X-ray spectroscopy when dealing with data reduction, best-fitting models and spectral simulations.

\begin{figure}
\includegraphics[width=0.6\textwidth]{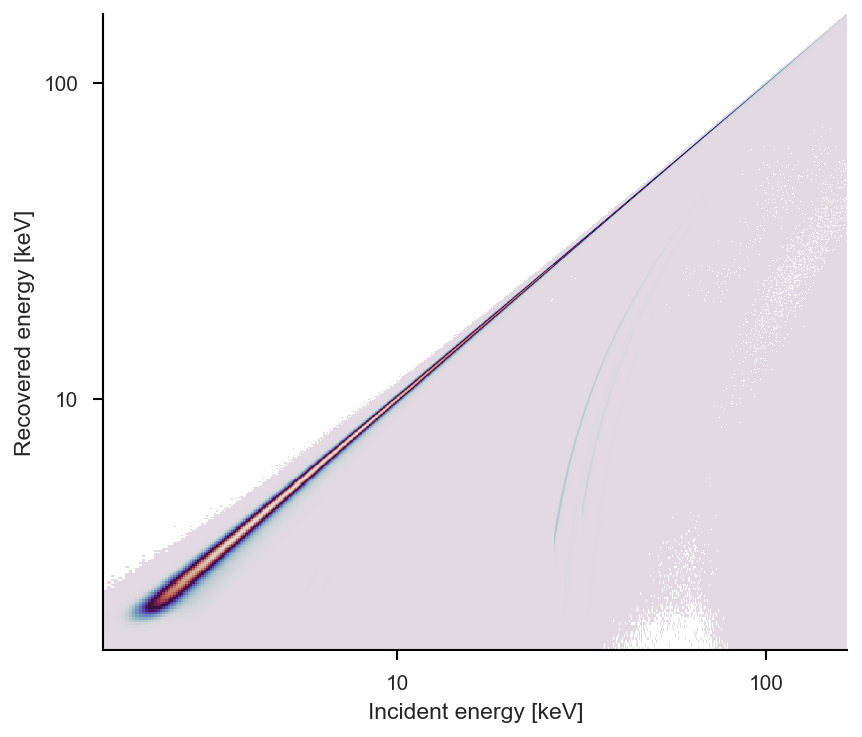}\label{fig:RMF}
\centering
\caption{Example of response matrix for  \emph{NuSTAR}/FPMB. Colored areas mark higher probability to recover a given incident photon energy, corrected for the energy-dependent effective area of the detector.
\label{fig:rmf}}
\end{figure}

\section{X-ray spectra and spectral modeling}\label{sec:models}

X-ray spectra represent the distribution of the observed emission as a function of energy. Describing this distribution through physical or phenomenological models can give insights on the physical processes at work in the observed source that are responsible for the observed emission.
Typically, an X-ray spectrometer collects event counts $C$ within each channel ($I$, which is specific for each telescope/instrument). The relationship between the observed spectrum $C(I)$ and the source spectrum $S(E)$ (where $E$ is the energy) can be described by:
\begin{equation} \label{eq. CI}
    C(I) = \int S(E) R(I,E) A(E) dE
\end{equation}
where $A(E)$ is the effective area of the instrument and $R(I,E)$ is the instrumental response (representing the probability of a detected photon of energy $E$ being registered in channel $I$, see Fig.~\ref{fig:rmf}).
The instrument response of an X-ray detector is affected by various physical constraints which limit its performance and introduce uncertainties in the observed spectrum.
As can be seen in Fig. \ref{fig:rmf}, there are two main structures that can be immediately identified at a glance in a response matrix: the diagonal and the off-diagonal parts.
Broadly speaking, the diagonal part describes the efficiency of the detector in accurately measuring the energy of the incoming X-rays. The off-diagonal part accounts for events where an X-ray photon of one energy is detected as having a different energy. It quantifies how much the energy of an X-ray photon is smeared in the detection process.
The diagonal and off-diagonal parts of the instrument response matrix have significant effects on the observed X-ray spectrum obtained from an astrophysical source.
These effects can include: energy calibration, energy resolution, line broadening, and spectral distortion (see Sect. \ref{section:low-high-res-spectra} and \citealt[][ and references therein]{Arnaud11}).

In principle, the more straightforward approach to find $S(E)$ would be by inverting Eq.~\ref{eq. CI}. However, this method leads to results which are highly dominated by noise \citep{Blisset79, Kahn80, Loredo89, Bouchet95}.
Therefore, the standard approach to fit data with physical models is by \textit{forward-folding}. 
In this method, a model spectrum, such as a power law, a blackbody, or any other type of phenomenological or physical model (based on physical considerations of the analyzed source)
is assumed and then convolved with the instrumental response.
This process returns counts in the instrument energy space, instead of the true photon flux spectrum emitted by the source.
The resulting model is then compared to the observed spectrum by means of an appropriate statistic (see \citealt{Buchner+Boorman2022} for details), and the model parameters can be iteratively adjusted until the process converges to the best-fit model, correspondent to the lowest possible value of the employed statistic. 
When the best-fit result is still unacceptable, a modification of the assumed model is required.

\section{Data structure and formats} 

Unlike detectors at longer wavelengths, X-ray astronomy instruments can detect each single photon from the pointed field of view, given the high energy budget of each event. 
Each X-ray photon is therefore characterized by a set of properties and other accessory information, most notably charge deposited as result of interaction with the detector which is related to photons energy, position in the detector plane, time of arrival, and recently also polarization. 
Detection of each photons is usually referred to as an ``event''. 
Note that wording event and photon are often used interchangeably as properties of an event that can normally be directly related to properties of a photon. Nonetheless, there are some differences which will be discussed later.
Depending on the detector properties,
additional information that may be useful for the screening and data reduction steps are included.
A list of detected X-ray photons constitutes the basic data structure in X-ray astronomy, and it is called an event file. 
The data in event files is typically derived from raw telemetry data and stored according to the Flexible Image Transport System (FITS) format\footnote{The International Astronomical Unit FITS Working Group: \url{https://fits.gsfc.nasa.gov/fits_standard.html}, \citet{Wells81}}, consisting of at least one Header and Data Unit (HDU, see, e.g.,  Table~\ref{table:header}).
To guarantee a consistent and formalized FITS files format, the Office of Guest Investigator Programs (OGIP) has organized a FITS Working Group (OFWG) to ensure and encourage that FITS files follow the same standard across the whole X-ray astronomy community \citep{Wells81}.

Among photon properties stored in event files, is their energy budget.
Usually, the deposited energy is detected as an electrical signal known as Pulse Height Amplitude (PHA).
PHA is proportional to the true photon energy through a constant known as ``gain'', which depends on energy, detector location, and is subject to shifts with time.
A reliable knowledge of the gain behaviour is therefore crucial for a correct instrument calibration and, for this purpose, monoenergetic radioactive sources are often equipped on board.
The gain corrected PHA, known as Pulse Invariant (PI), is sometimes preferred as the final product for spectral fitting. 
For the spectral data analysis, two additional important pieces of information are needed.
They are the Redistribution Matrix File (RMF) and the Auxiliary Response File (ARF). They are extensively described in another chapter of this book. Here, we just recall that the RMF provides the distribution of pulse height values for an incident photon of a given energy. High resolution instruments have an almost diagonal matrix, while proportional counters show non-zero matrix elements over a wide area of the matrix (Fig. \ref{fig:rmf}). ARFs contain important information like the effective area of the instrument and the quantum efficiency as a function of the energy. For some telescopes, a combination of the RMF and ARF is provided as a single file called ReSPonse Matrix File (RSP). 
This format is more compact and is obtained multiplying the ARF with the respective RMF. The calibration tool \texttt{rsp2rmfarf}\footnote{\url{https://heasarc.gsfc.nasa.gov/lheasoft/help/ftrsp2rmfarf.html}.} is available to split an RSP file back into a separate RMF and ARF.
In literature, RMF is sometimes referred to as ``response file''.

\begin{table}[!t]
\caption{A typical section of a FITS header storing information for a NICER simulated spectrum.}
\label{table:header}
\hskip-0.5cm
\begin{center}
\begin{tabular}{l c c c}
 \hline
 Attribute & Value & Comments \\[0.5ex]
  %& (Start) & [ks] \\[0.5ex] 
TTYPE1  = &CHANNEL            &  label for field   1\\
TTYPE2  = &COUNTS             &  label for field   2\\
TTYPE3  = &QUALITY            &  label for field   3\\
TTYPE4  = &GROUPING           &  label for field   4\\
...\\
TELESCOP = &NICER             &  mission/satellite name\\
INSTRUME = &XTI               &  instrument/detector name\\
CHANTYPE = &PI                &  channel type (PHA, PI etc)\\
...\\
RESPFILE = &nixtirefv001.rmf & redistrib matrix filename\\
ANCRFILE = &nixtiaveonaxisv005.arf           &  ancillary response filename\\
BACKFILE = &nixtiback20190807.pi           &  background filename\\
EXPOSURE =  &           2000. &  exposure (in seconds)\\
AREASCAL =  &                 1. &  area scaling factor\\
BACKSCAL =  &                 1. &  background file scaling factor\\[1ex]
 \hline
\end{tabular}
\end{center}
\end{table}

\section{Data reduction} 

To achieve reliable spectral analysis, it is essential to first perform the most appropriate data reduction. Afterwards, the spectra can be properly analyzed, that is modeled taking into account instrumental response, background, and statistical limitations of the available data. Here we describe the most common screening and reduction procedures to extract the events of interests, with emphasis on some common problems and pitfalls.

As already mentioned, the presence of background (noise) events in the data and others physical constraints of the X-ray detectors which affect the instrument response complicate the spectral analysis, making direct inversion of instrumental response unfeasible (which is the reason why the forward-folding approach is commonly adopted). Reducing background is, in fact, an essential step in any scientific analysis and X-ray spectroscopy is no exception. This task can be accomplished by improving signal to noise ratio through the use of X-ray focusing optics, passive and active shielding of the detector and using our knowledge of the detector properties to impose additional selection criteria (cuts) on the data. Some of the most common ones are discussed below.

\subsection{Pattern/grade selection}

\begin{figure}
\begin{center}
\includegraphics[width=\columnwidth]{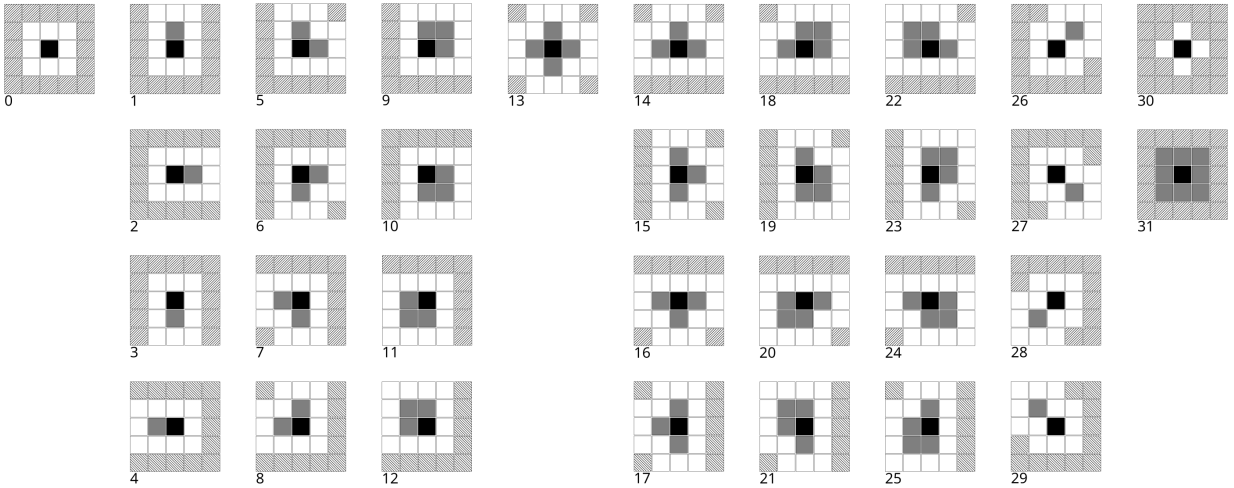}
\end{center}
\caption{EPIC MOS patterns (imaging mode). Each pattern consists of a $5\times5$ matrix centered on the pixel with highest charge (black pixel). Other pixels above the threshold are shown in gray, while those white are below the threshold. Pixels shown with lines-pattern can be below or above the threshold. For MOS, only patterns in the range 0$-$12 are used for scientific analyses.}
\label{fig:pattern}
\end{figure}

Many X-ray telescopes employ Charge-Coupled Devices (CCDs) as detectors.
Photons hitting the CCDs pixels release a charge proportional to their energy in the activated pixel and its neighbors. 
The charge cloud produced by an X-ray event and by charged particles hitting the detector can be confined to a single pixel or can be split over many pixels, depending on the cloud size relative to that of pixels. 
Each event is assigned a grade (or pattern) identification number, which describes the pattern of pixels whose charges exceed a predetermined threshold.
The pattern recognition scheme and combination of patterns calibrated depends on the telescope and detector.
For example, for EPIC/MOS the on-board software for pattern identification looks for signal enhancement (above a given threshold) in $5\times5$ pixel matrices scanned over the whole image.
Patterns unambiguously produced by non X-ray photons, such as cosmic-ray tracks produced by charge-particle background, are rejected.
Pattern selection can also be used to discriminate between true X-ray events with different quality:
the energy of the original X-ray photon is given by the sum of the charge over all the pixels interested by the event. Since each pixel includes some noise, X-ray events which split over many pixels have a relatively lower spectral resolution. Therefore, a set of valid patterns are selected to have a good trade off between quantity of the collected X-ray photons and quality of their spectral resolution.
For example, MOS in imaging mode has 32 predefined patterns, and those between 0 and 12 are considered ``valid''
(see Fig. \ref{fig:pattern}).
The grade is an important quantity that helps filtering out non-source events and to discard irrelevant events from on-board limited telemetry. 
It is also worth mentioning that the adopted pattern selection also affects the instrumental response, thus appropriate response files need to be used.

    \subsection{Cuts based on the background} \label{sect. bkg}
\begin{figure}
\begin{center}
\includegraphics[width=0.65\textwidth]{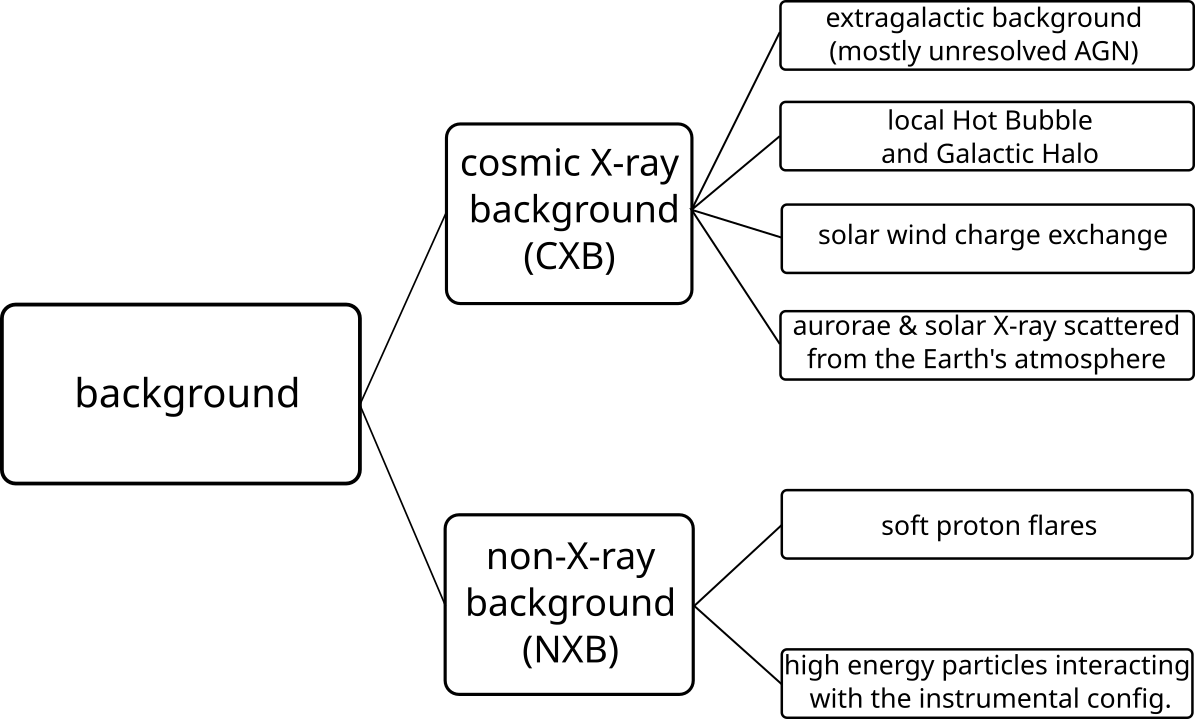}
\end{center}
\caption{Diagram showing the main different components of the X-ray background which can be detected by an X-ray telescope.}
\label{fig:bkg_scheme}
\end{figure}

Spectral analysis requires to take into account the contribution from the background,
which can be cosmic X-ray background (CXB) or particle-induced background.
CXB, in turn, is made of several components (Fig. \ref{fig:bkg_scheme}):
\begin{itemize}
    \item extragalactic background, mostly due to unresolved active galactic nuclei;
    \item Local Hot Bubble and Galactic halo;
    \item solar wind charge exchange, observed by \rosat\ and \xmm\ \citep[][ and references therein]{Freyberg20};
    \item aurorae and solar X-rays scattered from the atmosphere of the Earth.
\end{itemize}
As such, the CXB is an inherent component of sky emission and therefore can not be reduced, but only modelled.
The ``non-X-ray background'' (NXB), to differentiate it from the CXB,
is induced by particles and it is schematically divided in two major components:
\begin{itemize}
    \item secondary particles and X-ray photons produced by the interaction of cosmic rays (proton, electrons, He, with energies of $\gtrsim 100$~MeV) with the detector or with material surrounding it. The spectrum of this component varies with time and it is typically made of a continuum component produced by the interaction of the cosmic rays with the detector, X-ray continuum and fluorescence lines from the X-ray produced by the cosmic rays which hit the camera and the surrounding material;
    \item soft protons, likely organised in clouds populating the magnetosphere of the Earth, having energies of the order of $\approx 100-1000$~keV.
Their flux is modulated by the magnetic field of the Earth and strongly depends on the the time of the observation, location of the instrument and its pointing direction. These events are thus extremely variable (timescale of seconds-hours, with an increase of the background intensity by a factor of 10-10$^4$; \citealt{Kuntz08, Gastaldello17}). Background induced by soft protons affects X-ray telescopes with mirrors, like \emph{XMM-Newton} and \emph{Chandra}. In the lightcurves, they show characteristic bright flares that allows to remove the brightest events \citep{DeLuca04}.
\end{itemize}
The NXB contribution is effectively reduced through a proper design of X-ray telescopes, but some contribution always remains.
It is fundamental to accurately characterize the contribution from any cosmic and non-cosmic contribution to the observed spectrum.
The CXB contribution is usually estimated using an off-source region, and is either subtracted directly or included as part of the model  describing the overall spectrum.
Modelling of the NXB is generally more complicated.
The particle background has a different instrumental response than the CXB (NXB events come from all directions and some of them are not photons) and needs complex analysis of observations like, for example, those carried out with the closed filter wheel (e.g. \xmm; \citealt{Bulbul20}) or pointing to the night side of the Earth (e.g. \suzaku; \citealt{Yamaguchi06}). 
The observed NXB properties are then often correlated with other tracers to ensure appropriate scaling between calibration and actual observations.
    
In case of CCDs, dark currents can produce high noise, especially below $\sim 300$~eV. This may result in non-negligible events in some pixels (and sometimes in entire columns of the detector) where degradation occurs due to radiation or to manufacturing defects. Signals from those pixels should not be used in the spectral analysis and are usually excluded (flagged as bright and dead pixels). Note that exclusion of any detector parts also affects instrumental response and needs to be taken into account (i.e., the effective area is often calculated for each individual observation to account for excluded pixels and, therefore, reduced effective area).

\subsection{Pile-up and optical loading}

\begin{figure}
\begin{center}
\includegraphics[width=11cm]{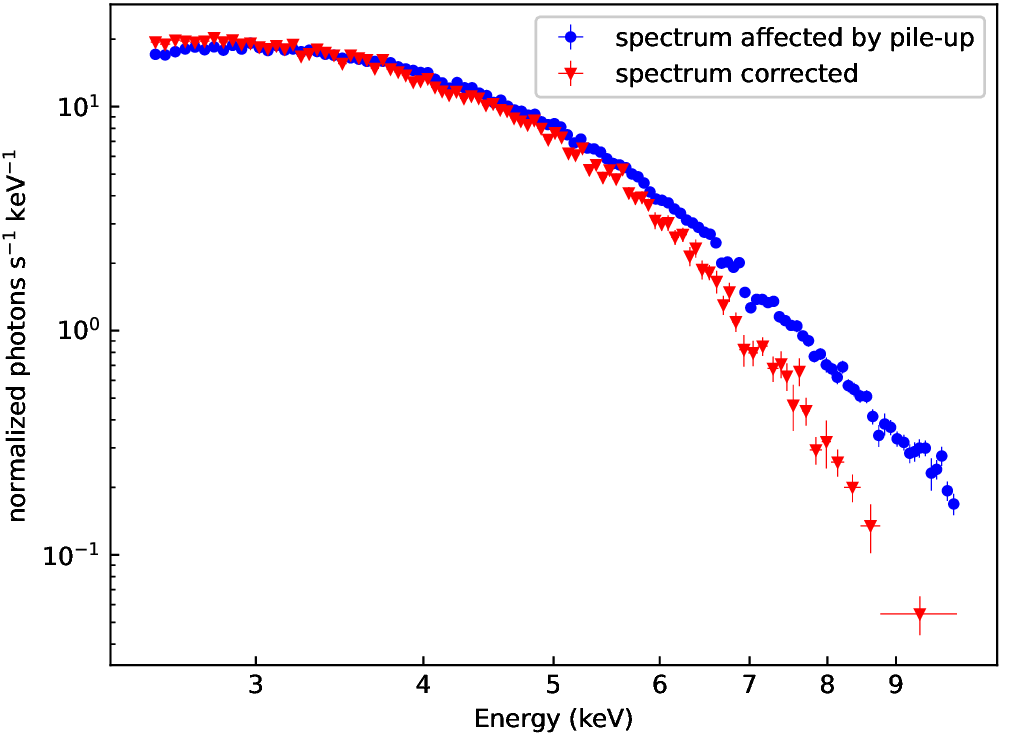}
\end{center}
\caption{EPIC MOS1 spectrum (imaging mode, 2.5$-$10~keV) of the low-mass X-ray binary GX~13+1 that shows the hardening of the spectrum due to pile-up (blue points), and the spectrum after correction (red points).}\label{gx13p1}
\end{figure}

    CCDs with relatively long read-out can be affected by the so-called photon pile-up when observing bright sources. 
    If two or more photons hit the same or adjacent pixels of the detector within one readout cycle, they are interpreted as a single event with energy equal to the sum of the energies of all the involved photons, which leads to distortions of the true spectrum.  
    
\begin{figure}
\begin{center}
\includegraphics[width=8cm]{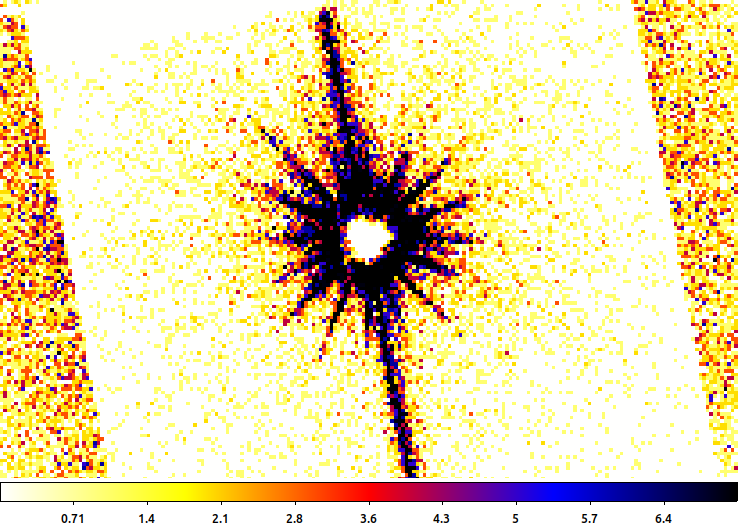}
\end{center}
\caption{EPIC MOS2 observation (imaging mode, 0.3$-$10~keV) of the low-mass X-ray binary Sco~X$-$1 showing a case of strong pile-up.}\label{fig:pileup}
\end{figure}

    The spectral shape is distorted by pile-up in three ways:
    \emph{i)} suppressing flux because of creation of invalid patterns; 
    \emph{ii)} by joining separate mono-pixel events into a single multi-pixel event (called pattern migration);
    \emph{iii)} hardening: the spectral response is compromised because the charge deposited by more than one photon is added up before being read out, creating artificial ``hard'' photons where there have actually been two or more soft photons (see Fig. \ref{gx13p1}). The Point-Spread Function (PSF) may be influenced by this effect: in its core, many photons arrive within one readout frame, creating multi-pixel photon patterns which are usually rejected by the onboard software (see Fig. \ref{fig:pileup}).
    
    Software for data analysis provide diagnostic tools, methods, or even corrections to the response matrix that can be applied to correct for pile-up, although full correction is not always possible. Therefore, when possible, the best strategy is to avoid pile-up entirely by appropriate planning of the observations.
    This can be accomplished by reducing the read-out time through selection of the operating mode for a given detector (i.e., by only reading part of the detector at higher speed), use of blocking filters or even opting for a more suitable telescope.

\begin{figure}
\begin{center}
\includegraphics[width=8cm]{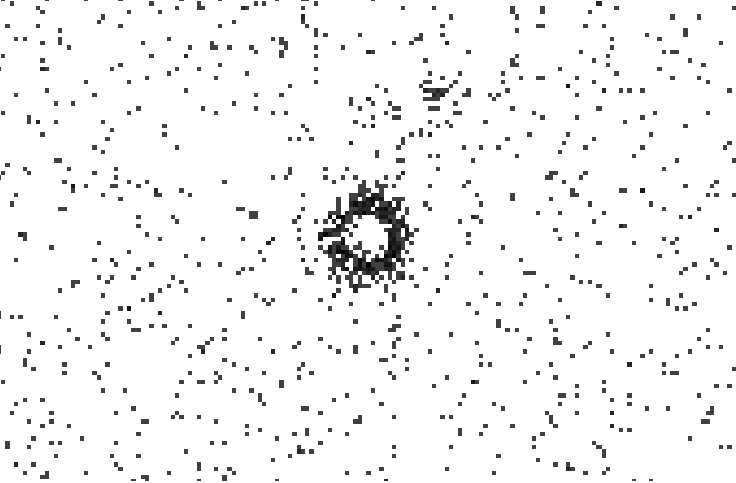}
\end{center}
\caption{Example of optical loading in a \emph{Swift}/XRT observation (0.3$-$10~keV). The star producing the ring feature, HD 46116, is a \mbox{V-mag$\approx 5.4$} star.}
\label{fig:optical_loading}
\end{figure}

An effect similar to pile-up is optical loading. X-ray CCDs are usually equipped with optical blocking filters, to reduce the amount of optical photons (ultraviolet, optical, infrared) that reach the CCD. Nevertheless, particularly bright optical sources can still cause significant charge in the CCDs, when the number of electrons liberated by optical photons in a CCD pixel between successive read-outs  is large enough to be erroneously classified as an X-ray photon. This anomaly is known as optical loading. It increases the local background, leads to false detection of X-ray sources, and for particularly bright sources, it can affect the X-ray event energy calibration and grade, leading to spurious ring-like structures in the images (see Fig. \ref{fig:optical_loading}).

    \subsection{Selecting events of interests}
    
    Events of interest are considered valid when they pass certain discriminatory criteria. To this end, Good Time Intervals (GTIs) are identified by filtering the observation in order to remove high-background episodes, such as SAA passages for low-Earth orbits, rejection of events from known hot pixels, etc.
    This results in cleaned events that are successively calibrated according to the instrument specifics.
    For non-imaging X-ray instruments, the source of interest consists of all cleaned and calibrated events in the instrument field of view.
    However, for imaging X-ray telescopes, a source and a background region can be distinguished on the detector, and their events can be extracted separately.
    There is a third option, consisting of coded mask telescopes, for which each source in the field of view contributes to the background of all other sources as well.   
    
    For imaging detectors, to maximize the signal-to-noise ratio of the source counts with respect to the background level, assuming a point-spread function, some software tools are available to calculate the optimal centroid and size of the source extraction region (circle or ellipse), such as, for example, \texttt{eregionanalyse} for \emph{XMM-Newton}/EPIC data\footnote{\url{https://xmm-tools.cosmos.esa.int/external/sas/current/doc/eregionanalyse/index.html}} and \texttt{nustar-gen-utils} for \nustar\ data\footnote{\url{https://github.com/NuSTAR/nustar-gen-utils/blob/main/notebooks/OptimalRadius_Example.ipynb}}.
    Some rules of thumbs to determine reasonably good shapes and sizes for the extraction regions of source and background are the following:
    \begin{itemize}
    \item background extraction area larger than that of the source. Background level is expected to be much lower than that of the source. To increase the statistics and have a better modeling of the background, it is recommended, whenever possible, to have a relatively wide area to extract the background spectrum.
    \item background extraction radius should not be contaminated by the target itself and by any other source;
    \item background extraction area should not be located too far from the source, to have similar instrumental properties of those of the source (otherwise, response file should be calculated also for the background). Usually, this leads to choose whenever possible an annulus region, with, for example, inner radius $\sim 2\times$ the extraction radius used for the source, and $\sim 3\times$ for the outer radius.
    \item Differences in area between source and background have to be taken into account for the spectral analysis by appropriate rescaling.
    \end{itemize}
    Typically, the background extraction region area is taken into account by the data reduction software through a keyword in the FITS header (\texttt{backscale}) that scales its size to that of the source.
    For non imaging detectors, background spectrum is calculated using methods developed by the instrument team (see Sect. \ref{sect: Background treatment}).
    
    To study the spectral evolution of a source with a periodic signal (for example, the spin period of a pulsar/white dwarf, orbital and superorbital periods) or the spectral variability as function of the
    total flux of a source or any other physical quantity, appropriate good-time-intervals (GTIs) have to
    be created and applied to extract spectra within the desired time intervals.
    In this framework, the most common spectral analyses involve data extraction within phase intervals and it is often called ``pulse phase resolved spectroscopy", while that involving spectral extraction within some flux intervals is called ``flux resolved spectroscopy''.
    
    Some detectors offer data acquisition modes in which a spatial dimension is collapsed to enable a faster read out by the electronic system. These are, for example the ``timing'' mode in \xmm\ and the ``window timing'' mode in \swift. A trickier situation is when the data come from a grating spectrometer, like the Reflection Grating Spectrometer on board \xmm, and the spectrometers on board \chandra. In these cases, the best methods to select source and background regions in which the spectra have to be accumulated are not trivial. Thus, they are provided or made more automated within the data reduction software by the instrument and software teams.

\section{Software for spectral analysis}

There are several software packages aimed to simplify analysis of X-ray spectra.
These software are generally able to handle an instrumental response, provide several types of spectral models, as well as statistical and visualization tools.
  Their ultimate goal is to facilitate the analysis and interpretation of a broad variety of spectral properties from many different types of X-ray sources, to facilitate the understanding of the origin of their X-ray emission.
  
  One of the most used software for spectral analysis is XSPEC\footnote{\url{https://heasarc.gsfc.nasa.gov/docs/xanadu/xspec/}} \citep{Arnaud96, Dorman01}. 
  It is mainly intended for interactive work but also has different programming languages (such as Python and TCL) interfaces for scripting.
It also benefits from Heasp\footnote{\url{https://heasarc.gsfc.nasa.gov/docs/software/lheasoft/headas/heasp/heasp_guide.html}}, a C/C++/Python library whose aim is to provide tools to handle PHA, RMF, ARF, and XSPEC table model files.
  
  \emph{Sherpa}\footnote{\url{https://cxc.cfa.harvard.edu/sherpa/}} is the Python modeling and fitting application by the \chandra\ X-ray Center (CXC), provided within the \chandra\ Interactive Analysis of Observations (CIAO) software package \citep{Freeman01, Doe07}. It can be used for many types of analyses (spectral, images, timing) and telescopes, including those operating in X-ray band. 
  Sherpa also includes most of the XSPEC models.
  
  The Interactive Spectral Interpretation System (ISIS)\footnote{\url{https://space.mit.edu/CXC/ISIS/}} is based on S-Lang scripting interface and is especially designed for the analysis of high resolution X-ray spectra \citep{Houck00}.
  The strength of ISIS and Sherpa relies in the fact that they have been developed to ease scripting implementation.
  
  SPEX\footnote{\url{https://www.sron.nl/astrophysics-spex}} is a software package provided by SRON Netherlands Institute for Space Research optimized for the analysis of high-resolution X-ray spectra. Similarly to XSPEC, it is accompanied by a Python toolbox, called Pyspextools \citep{Kaastra96,Kaastra20}.
  
  Line identification is an important part of the spectral analysis, especially if dealing with high-resolution spectra (see Sect. \ref{section:low-high-res-spectra}). This is usually accomplished through the use of three atomic databases: AtomDB, SPEX, and CHIANTI. AtomDB and SPEX focus more on the X-ray ranges, while CHIANTI has its main focus on far-UV/soft X-rays \citep[see, e.g., ][]{Foster12}.
  There are a number of software packages that provide several tools which enable the users to perform interactive analysis of line emission (their identification, measure line fluxes, modeling) and to facilitate the comparison of the observational data with existing atomic line databases.
  Some of them are PyAtomDB and ChiantyPy, based on Python, and PintofALE, based on IDL.

\section{Spectral analysis}
\label{sect:spectral analysis}

\subsection{How to fit and how to test a spectral model} \label{sect:how}

\begin{figure}
\begin{center}
\includegraphics[width=0.65\textwidth]{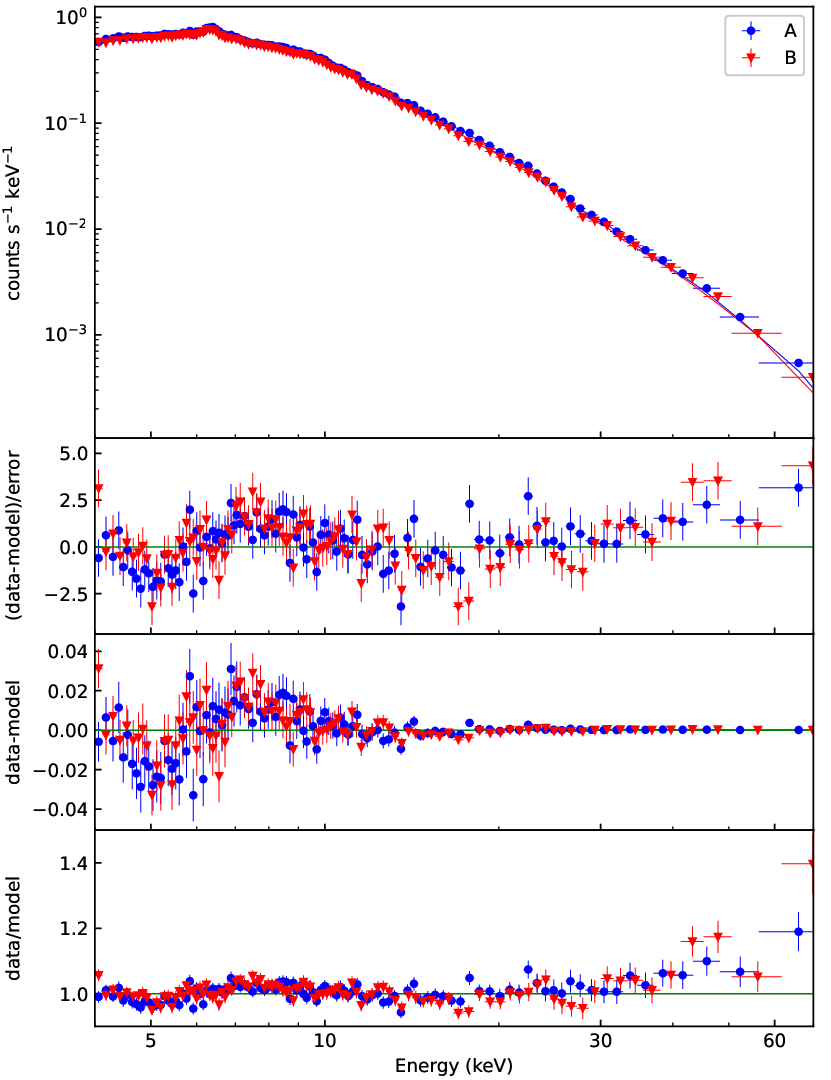}
\end{center}
\caption{\emph{Top panel:} example of a \emph{NuSTAR} count spectrum (A and B: spectrum from module FPMA and FPMB, respectively), fitted with an absorbed power law with high energy cutoff and a Gaussian line at energy $\sim 6.4$~keV. \emph{Three lower panels:} various types of residuals defined in XSPEC.}\label{fig:residuals}
\end{figure}  

Forward-folding approach (see Sect.~\ref{sec:models}) works through the iterative minimization of a given \textit{fit statistic} to find the model best-fit parameters and employs a certain \textit{test statistic} to test the best-fit model.

Spectral models can be defined by the user as a mixture of individual components. 
For practical reasons, software packages such as XSPEC differentiate between separate components categories, namely \textit{additive, multiplicative, convolution}, and \textit{mixing} models.
Additive models represent emission mechanisms (e.g., an emission Gaussian line, a blackbody, or a power law component), while multiplicative models are energy-dependent modifications of the continuum emission (e.g., photoelectric absorption, high-energy cutoff, or Gaussian absorption lines).
Additive and multiplicative models can be either directly coded or implemented as so-called \textit{table models}.
These are grids of spectra tabulated for different values of the parameters of the model. 
They can be additive, multiplicative, and exponential. The correct model spectrum for a parameter value is obtained by linear interpolation on the grid. 
Plasma codes like XSTAR \citep{Kallman01}, CLOUDY \citep{Ferland17}, Titan \citep{Dumont00}, Mocassin \citep{Ercolano03}, are often used to generate grids of models which are then stored in table model FITS files.

Many software packages offer the possibility to create user-defined models. These models can be defined by the user as arithmetic expressions. For example, the tool \texttt{mdefine} in XSPEC allows the user to define relatively simple functions. Additional publicly released models have been made available by their authors\footnote{\url{https://heasarc.gsfc.nasa.gov/docs/xanadu/xspec/newmodels.html}.}.
Moreover, to facilitate the software usage, models and data as well as general environment settings, can also be uploaded in XSPEC through \texttt{xcm} command text files (see Sect.~\ref{sec:performance}).

While multiplicative components apply a factor that depends on the energy, another class of models, called ``convolution'', modify the underlying model through more complicated transformations performed using a convolution operator. The most common convolution models across the different software platforms are, for example, those that perform smoothing with some function (e.g. Gaussian, Lorentzian), Compton reflection, and redshift of an additive component. Some particular convolution models allow to calculate flux, luminosity, and even correct for pile-up.
Mixing models perform complicated ``mixing'' transformations on spectra from different data groups. In many cases, additional information have to be included in the input spectra to apply on them mixing models, and they are usually located in the FITS
XLFT keywords. They are used for a number of tasks,
such as to estimate the surface brightness, the distribution of gravitating mass in atmospheres (the so called ``cluster mass mixing models''), and to obtain 2-D  projected spectra from 3-D regions (see {\tt projct} model in XSPEC). 

According to the quality of data, different test statistics need to be considered in order to assess if the employed model properly fits the data.
For sources with count rates high enough to consider the distribution of counts in each energy channel to be Gaussian, a $\chi^2$ statistic is typically considered.
In this case, a widespread, although not always statistically sound practice, is that to regard as acceptable a model when the ratio between the $\chi^2$ test statistic value and the degrees of freedom -- i.e., the reduced $\chi^2$ ($\chi^2_{\rm red}$) -- is approaching unity. If $\chi^2_{\rm red}$ is much larger than unity (e.g., $\chi^2_{\rm red}>1.2$) it likely implies that errors on data are under-estimated and/or the model does not describe the data well enough. On the other hand, if $\chi^2_{\rm red}$ is much smaller than unity (e.g., $\chi^2_{\rm red}<0.9$) it likely implies that errors on data are over-estimated (e.g., employing a systematic error that is too large), and/or that the employed model is overparameterized.
However, such a usage of the $\chi^2$ test statistic is often unjustified and should be applied with care \citep[see, e.g., ][]{Andrae2010, Kaastra17}.

When performing X-ray spectral analysis, it is common practice to display a plot consisting of an upper panel with the observed X-ray spectrum and its best-fit model, and a lower panel with the so-called ``residuals''.
Both panels provide a visual assessment of the goodness of fit between a given model and the observed data.
Residuals are usually defined in three different ways:
\begin{itemize}
\item \emph{data-model:} differences between the observed data and model predicted data (keyword: \texttt{residuals} in XSPEC);
\item \emph{(data-model)/error:} residuals are displayed in terms of the sigmas (keyword: \texttt{delchi} in XSPEC). In the XSPEC implementation, for Cash and $W$ statistic, the errors of the residuals are calculated as the square root of the number of counts predicted  by the model;
\item \emph{data/model:} residuals show the data divided by the model (keyword: \texttt{ratio} in XSPEC).
\end{itemize}
The objective of viewing the residuals panel is to assess the overall quality of the model fit. A good fit should have residuals randomly distributed around zero with no systematic trends or large discrepancies. If significant patterns or structures are present in the residuals, this suggests that the model is not an adequate representation of the true underlying emission process.
Significant positive or negative residuals indicate regions where the model under- or over-predicts the observed flux. These regions may be associated with specific emission or absorption lines, a poor choice of the model components used to describe the continuum (for example, because other physical processes need to be included in the model). Residuals can also be important for the identification and evaluation of possible systematic errors\footnote{See, for example, \href{https://heasarc.gsfc.nasa.gov/docs/nicer/analysis_threads/plot-ratio/}{https://heasarc.gsfc.nasa.gov/docs/nicer/analysis\_threads/plot-ratio/}} and to highlight instrumental artefacts.
A common approach to improving the model or gaining more insights into the physical processes involved is to iteratively adjust the model parameters and re-examine the residuals and the test statistic value (e.g. $\chi^2/$d.o.f.), until a satisfactory fit is achieved. The ultimate goal is to find the best-fitting model that accurately describes the observed X-ray spectrum, helping to infer the properties and characteristics of the astrophysical object under study.
The top panel of Fig. \ref{fig:residuals} shows an example of the plot typically used to display the observed X-ray spectrum and its best-fit model. The three lower panels show the three possible ways to define the residuals in XSPEC: \texttt{residuals}, \texttt{delchi}, \texttt{ratio}. The final choice of the type of residuals to be displayed is the responsibility of the scientist, whose ultimate aim is to display accurate residuals and a correct picture of the results of the spectral analysis.

Alternatively to $\chi^2$, a number of other test statistics are available and should be carefully considered according to each case.
For example, the Kolmogorov-Smirnov test has proved to be more advantageous with respect to the $\chi^2$ in many cases \citep{Lilliefors67}.
Once a test statistic has been applied, a goodness-of-fit test must be ran in order to assess if the null hypothesis that the observed data are drawn from the model can be rejected. This can be done by calibrating the goodness-of-fit statistic via bootstrap methods, e.g., using the \texttt{goodness} tool in XSPEC. Such tool builds a distribution of the test statistic by simulating data that are based on the best-fit model, thus allowing to compare the employed test statistic against the estimated distribution.

\begin{figure}
\begin{center}
\includegraphics[width=0.49\textwidth]{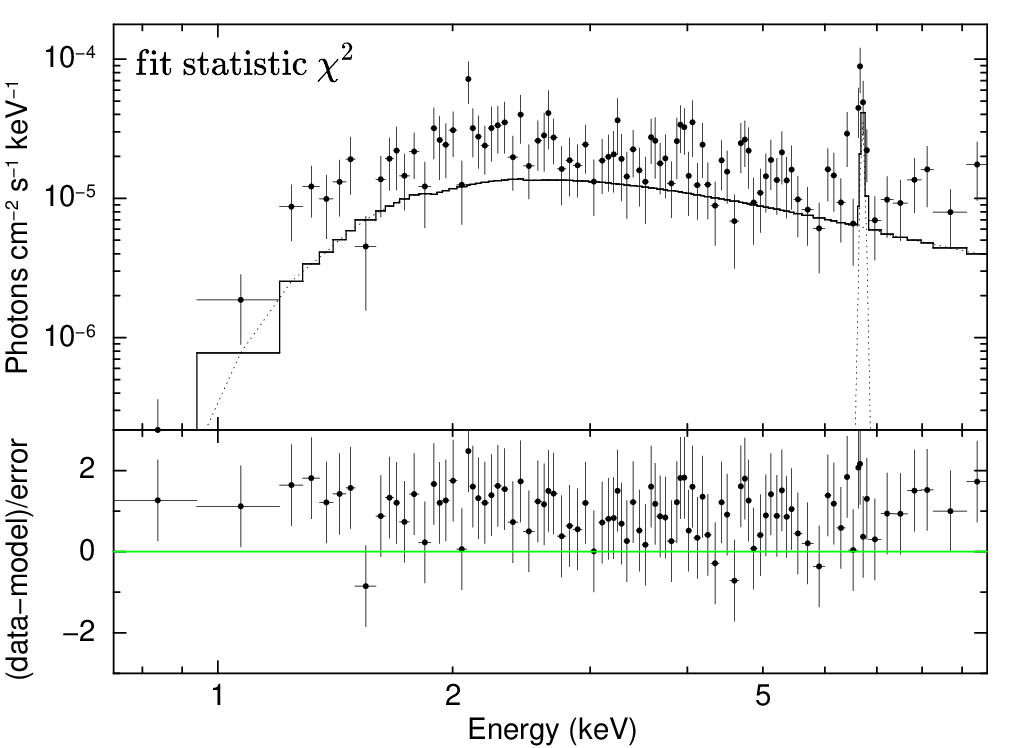}
\includegraphics[width=0.49\textwidth]{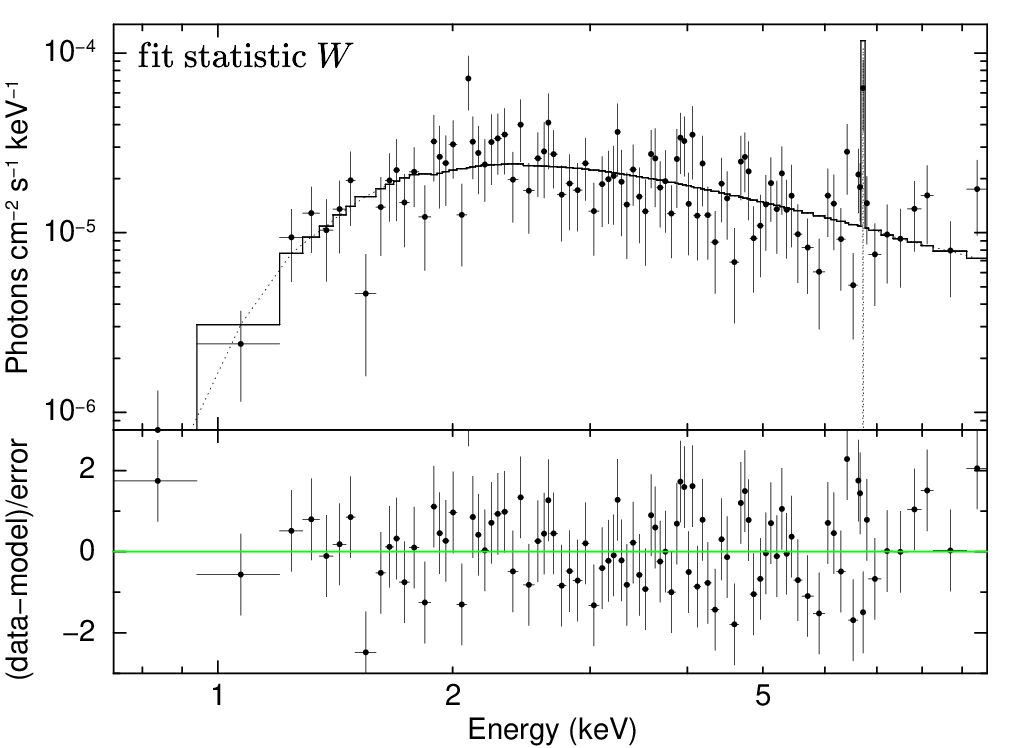}  
\end{center}
\caption{Example of an X-ray spectrum with small number of counts per bin (note that the spectrum shown here is rebinned for visual clarity). \emph{Left panel:} the spectrum is fitted using $\chi^2$ statistic: top panel shows the best-fit model (solid line), which is sistematically below the data points. \emph{Right panel:} the same spectrum is fitted using $W$ statistic. Bottom panels show the differences between the data and the model. In this example the source is Swift J173518.7$-$325428 as observed by \xmm\, (MOS1 data, imaging mode) on 31 March 2013.}\label{fig:chi2_vs_cstat}
\end{figure}

For a proper test statistic to work as an appropriate goodness-of-fit estimator, a suitable fit statistic also needs to be adopted.
In case of Gaussian distributed data, the $\chi^2$ statistic is often used to find both the best-fit model and its best-fit parameters, i.e., $\chi^2$ is employed both as a test statistic and as a fit statistic.
Using the proper fit statistic and the proper channels binning (either by a minimum count rate per bin or by optimal binning techniques -- e.g., \citealt{Kaastra+Bleeker16}) is essential for a proper spectral fitting procedure.
However, for a broad class of cases, adopting the $\chi^2$ statistic may lead to biased results \citep{Nousek89, Humphrey09}.
For example, when $\chi^2$ statistic is used with spectra having small numbers of counts per bins, the best fit model can lie below most of the data points and the fit statistic value could result suspiciously low (see left panel of Fig. \ref{fig:chi2_vs_cstat}).
This is caused by the incorrect use of $\chi^2$ statistic, being the data Poisson distributed. Data bins with counts lower than those expected by the model are weighted more than the other bins, favouring an overall lower weight for the best fit model.
To deal with these special cases, different methods have been proposed
to modify the $\chi^2$ fitting technique through a different weighting of each bin.
For example, \citet{Gehrels86} derived a different approximation for the standard deviation, $1+\sqrt{N+0.75}$ (where $N$ is the number of counts for a given spectral bin), used to weight each bin.
Another method proposes a smoothing of 
the neighbours channels using a sliding window technique described in \citet{Churazov96}. 
Despite those alternative methods, it is better considering a generally unbiased statistical approach altogether, such as the Cash fit statistic \citep{Cash79}.
This is typically the case for low counts spectra, where the energy channel distributions of counts are considered Poissonian, but it applies in the high count regime as well \citep{Humphrey09}.
In the case of a Poissonian background, the Cash fit statistic is modified in order to take into account the combined likelihood for source and background, also dubbed as the W statistic \citep{Wachter79}.
Moreover, software such as XSPEC can simultaneously fit different data groups using different fit statistics, to be consistent with different data sets.

\subsection{Spectral energy resolution and binning} \label{section:low-high-res-spectra}

X-ray telescopes like EPIC on board \xmm, \chandra, \erosita, use CCDs which provide at best spectral resolutions of the order of $R=E/\Delta E \approx 10-50$. This is enough to measure the main physical parameters from the continuum emission (for example, the absorption column density,
the temperature of a blackbody, bremsstrahlung, and the non-thermal emission) and, in many cases, 
to study the fluorescence emission line of Fe K$\alpha$, that is often observed in X-ray binaries, cataclysmic variables, and AGNs, as well as cyclotron resonance scattering features observed in many spectra of strongly magnetized accreting pulsars and magnetars \citep{Staubert19, Tiengo13}. Spectra with numerous lines from abundant elements (e.g. Fe, O, Ne, Mg, Si, S) 
are usually beyond the capabilities of these instruments. 
For example, one of the problems of low-resolution spectra is that they are more prone to line blending issues, which lead to wrong estimates of fit parameters.
Since the major atomic transitions of astrophysical plasmas appear as spectral features in the X-ray band, their study is of great importance, and higher spectral resolution is thus required. 

Ideally, high-resolution spectra should allow to resolve the shape of individual lines. In the $\sim 0.1-10$~keV this requires $R\approx 10^4$.
Diffraction gratings on board of \chandra\ and \xmm\ have spectral resolving power up to $\sim 300$. 
X-ray microcalorimeters offer a better spectral resolution, as evidenced by the results from the Soft X-ray spectrometer (SXS) on board the \hitomi\ satellite, which performed the first X-ray observations with a microcalorimeter in the energy band 0.3$-$12~keV. SXS had a resolving power of $R\gtrsim10^3$ above $\sim 4$~keV that enabled to obtain fundamental results about chemical abundances, previously unidentified line emissions and absorption, and to get fundamental clues on turbulences in the intracluster plasma.
It also provided an important test for theoretical models of X-ray emission from hot collisional plasmas \citep[][ and references therein]{Ezoe21} and references therein.
Two other missions operating in the same energy range as SXS are planned: XRISM and \emph{Athena}. They host a transition edge sensor (TES) microcalorimeter (for TES microcalorimeters see, e.g., \citealt{Gottardi21}), which is expected to achieve a resolution power of $R\approx 2400$ at $\sim 5.9$~keV.

It is also worth mentioning the spectrometer on board of INTEGRAL (SPI), a gamma-ray telescope that operates at the much higher energy range 20~keV to 8~MeV, compared to the instruments discussed above. Thanks to its high energy resolution ($\sim2.5$~keV at $\sim 1.3$~MeV), it is able to measure intensities, profiles, and shifts of gamma-ray lines, especially from nucleosynthesis and from other X-ray and gamma-ray point and extendend sources \citep{Vedrenne03}.

Not all targets are suitable for a given high-resolution X-ray spectrometer.
If we focus on the energy range 0.2$-$10~keV, slitless grating spectrometers have a good energy resolution below $\sim 2$~keV but suffer of degradation for extended objects and for point sources in a crowded region or embedded in a diffuse emission. X-ray microcalorimeters provide better spectroscopic capabilities compared to gratings, with an almost constant energy resolution as function of the energy (also above 2~keV). They are also more suitable to study extended sources (providing their spatial resolution is sufficient).

Data reduction of low- and high-resolution spectra depend on the properties of each instrument. For high-resolution spectra, response matrices may have very large sizes and the adopted spectral models can have many free parameters. This can lead to high computationally intensive fitting runs, which might be critical for many computer systems. To handle analysis of high-resolution spectra in a computational efficient way, expressions to calculate the optimal bin size for observed data and model spectra have been derived by \citet{Kaastra+Bleeker16} and are known under the name of ``optimal binning''.
Other binning algorithms and methods are available.
For example, the \texttt{ftgrouppha} within \texttt{heasoft} includes the \texttt{optimal} binning option along with the possibility to group the bins according to a minimum number of counts.
Another example is the \texttt{specgroup} tool within \texttt{XMM-Newton} Science Analysis System (SAS) software which allows, among other options, to copy the grouping from another spectrum or to rebin the spectrum such that each bin has a minimum ratio of source signal above background.
It is worth mentioning that most of those methods do not actually regroup the spectrum, but rather flag entries in the ``GROUPING'' column of the spectrum in a way that the software can recognize which bins belong to a given channel.
To actually modify the spectral binning users should use \texttt{ftrbnpha} along with the \texttt{ftrbnrmf} tools to rebin the RMF to obtain the desired number of channels and energy bins.
This last option, unlike grouping, is irreversible.
It is also important to recall that binning can be needed to enable the use of different fit statistics (for example, $\chi^2$ or Cash).

When dealing with spectra of sufficiently high spectral resolution and statistics, it may be possible to observe another relevant effect in X-ray spectroscopy, called ``escape peaks''. They are artificial peaks that appear in the spectrum. They occur when an incident X-ray photon interacts with the detector material and loses some of its energy, but does not fully deposit all its energy within the detector. The ``escaping'' photon, leaves behind a characteristic energy signature that corresponds to the remaining energy after the interaction, which  results in a spurious peak in the spectrum. Escape peaks can interfere with the analysis of the spectrum, as they may be mistaken for real peaks from the source or the material between the source and the observer \citep{Grimmm09, Arnaud11}.

\subsection{Background treatment}
\label{sect: Background treatment}

A proper treatment of the background spectrum is key for a consistent fitting procedure. 
When an X-ray source has enough counts per channel, or when channels are combined to contain each enough counts,
data errors are in good approximation Gaussian, and the $\chi^2$ fit statistic can be used. In this case, a background spectrum (extracted from a source free region, or calculated using other methods) can be subtracted from the source spectrum, and variances of source and background are added in quadrature. 
The subtraction of the background spectrum is allowed because the difference of two Gaussian distributions remains Gaussian. This is not the case for data that are Poisson distributed, i.e., in the low counts limit. In this case, as mentioned in Sect. \ref{sect:how}, if the data set containing the source spectrum is loaded with an associated data set for background component, W statistic should be used. Nonetheless, the correct behaviour of W statistic in regimes with very low counts is not always guaranteed. For example, it has been shown that when the background has many channels with zero counts, the fit can lead to incorrect results. One solution is to rebin the source and background spectra to have at least a few counts per bin in the background spectrum\footnote{See: \url{https://giacomov.github.io/Bias-in-profile-poisson-likelihood} and, for example, \citet{Snios20}.}. The downside of this approach is the loss of sensitivity to spectral features in the background narrower than the newly defined (i.e. larger) bins \citep{Kaastra+Bleeker16}.
An even better (but more complicated) approach is to fit the source and background spectra (loaded as different data sets) simultaneously with two separate models, one for the source, one for the background, using Cash statistic. This method also allows to use distinct RMFs and ARFs for source and background spectra (for example, to take into account the variability of the instrumental response along the detector, if the source and background extraction regions are far from each other\footnote{See the example in: \href{https://asd.gsfc.nasa.gov/XSPECwiki/background}{https://asd.gsfc.nasa.gov/XSPECwiki/background}}).
A handy example of background modeling is provided by the NICER SCORPEON tool\footnote{\url{https://heasarc.gsfc.nasa.gov/docs/nicer/analysis_threads/scorpeon-overview/}}.
As the NXB and CXB are variable in time and sky location, respectively, SCORPEON produces a background model specific for a given set of observations by taking into account the following components: South Atlantic Anomaly (S), Cosmic Rays (COR), Polar and Precipitating Electrons (PE), cOnstant terms (O), Noise peak (N).
Additional background terms, not mentioned in the acronym, are also included in the model.
Among the most important benefits of this approach, we highlight that modeling the background components likely returns a better fit of the source data and a more realistic estimate of the best-fit parameters.

\begin{figure}
\begin{center}
\includegraphics[width=0.27\textwidth]{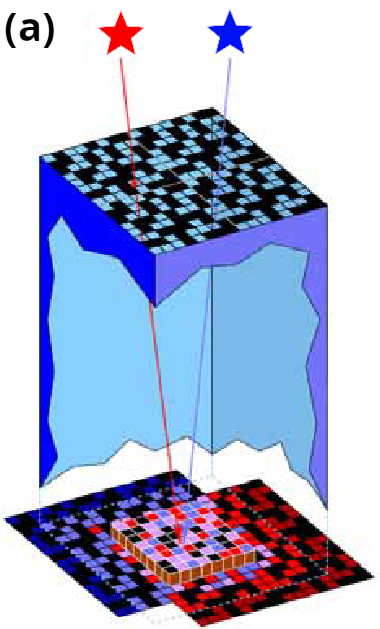}\hspace{0.4cm}
\includegraphics[width=0.63\textwidth]{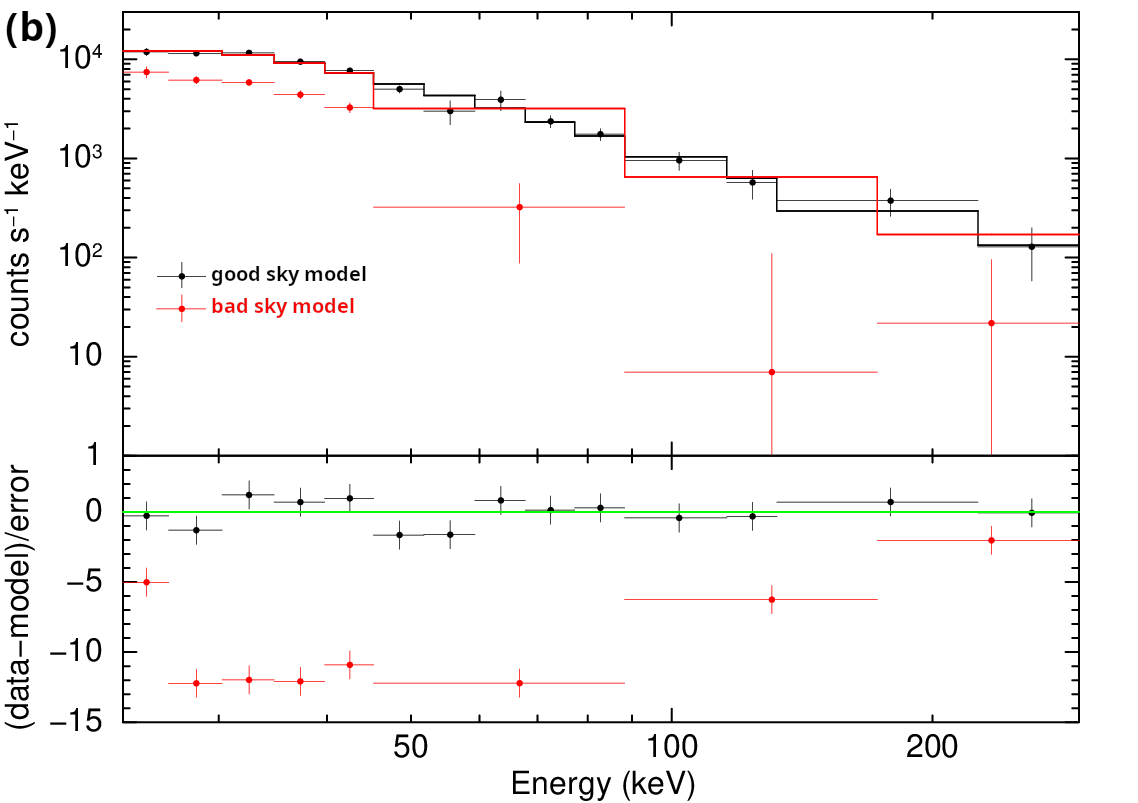}\\
\vspace{0.4cm}
\includegraphics[width=0.98\textwidth]{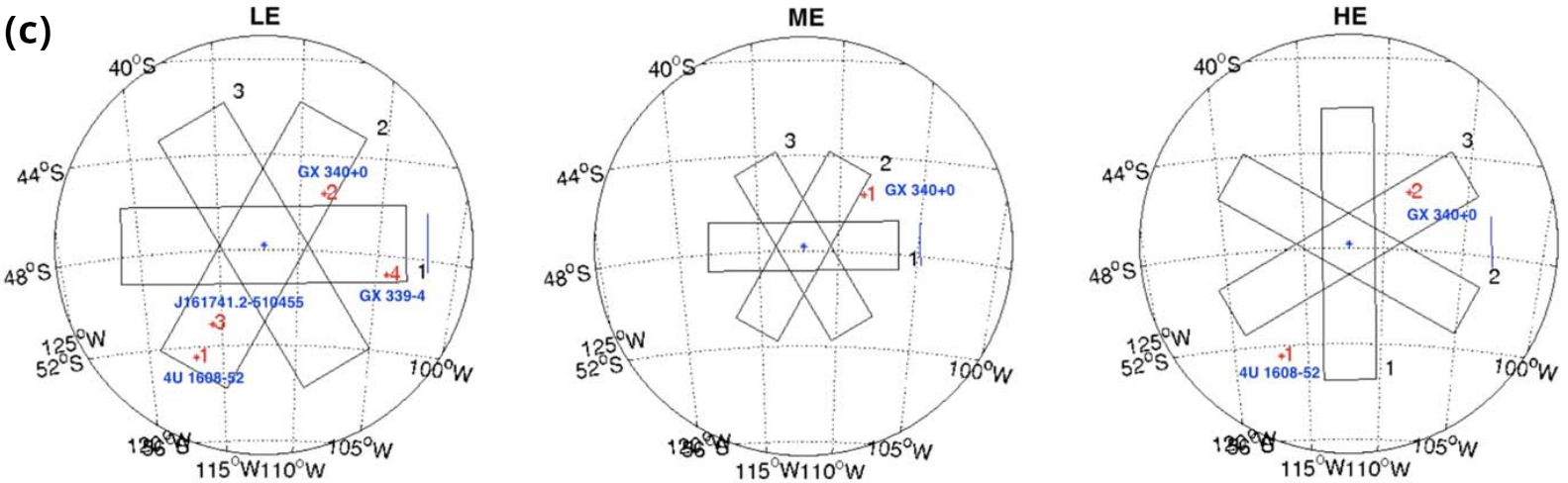}
\end{center}
\caption{Some examples of possible issues in the analysis of spectra from X-ray coded mask and collimator telescopes. \emph{Panel a:} in coded mask telescopes, the shadow of two or more sources in the field of view are projected on the detector. Each source is the background of all other sources in the field of view (Credits: ISDC/M.~T\"urler; \href{https://www.isdc.unige.ch/integral/gallery.cgi?ALL}{https://www.isdc.unige.ch/integral/gallery.cgi?ALL}). \emph{Panel b:} if the sky is not modeled properly, the spectrum of a source can be inaccurate. Here is a comparison between the INTEGRAL/SPI spectrum of Cyg~X$-$3 obtained assuming a correct sky model in which other bright sources in the field of view are considered, namely Cyg~X$-$1 and EXO~2030+375 (black points), and the other one (red points) obtained with a wrong sky model, in which Cyg~X$-$1 and EXO~2030+375 are not considered. The residuals displayed in the lower part of the plot show a significant departure of the red data from a model obtained with a more accurate sky modeling (green line). \emph{Panel c:} An example of strategy to account for contaminating sources in the spectrum of the accreting black-hole MAXI~J1631$-$479 observed by HXMT detectors LE, ME, HE. The detectors are composed by three Detection Boxes. Estimates of flux differences using lightcurves obtained from each of them allowed to estimate and remove the contribution from GX~340+0, which would otherwise have contaminated the spectrum of MAXI~J1631$-$479 \citep{Bu2021}; \copyright AAS. Reproduced with permission.}\label{fig:integral_hxmt}
\end{figure}

In observations performed with coded mask telescopes, in addition to the NXB and CXB, each source is background for the others (Fig. \ref{fig:integral_hxmt}, panel \emph{a}). Therefore, during the data reduction process, it is necessary to provide the software with an appropriate ``sky model''. The sky model contains information about the active sources in the field of view, their time variability, and the background properties (i.e., level of CXB and NXB background). The sky model is then used in a process called ``deconvolution'' to extract the correct flux of all the active sources. An incorrect sky model may lead to flawed (and in some cases largely wrong) results, as shown in Fig. \ref{fig:integral_hxmt} (panel \emph{b}).
Spectra obtained with other types of X-ray telescopes that use collimators might also be unreliable, due to contamination from other sources in the field of view. Depending on the properties of the instrument, some strategies can be adopted to account for contaminating sources effects in the spectra obtained by these telescopes (see the example shown in Fig. \ref{fig:integral_hxmt}, panel \emph{c}).

\subsection{Testing model components} \label{sect: testing model components}

Most often, the spectral models employed to fit the data consist of multiple components. To avoid overcomplicated models, it is crucial to assess the significance of a given component.
The way this is done depends on the spectral component type (additive, multiplicative, etc.) that needs to be tested.
For instance, a typical case is represented by the significativity of the Iron K$\alpha$ emission line at rest energy of 6.4 keV detected on top of a continuum model.
Although it has become common practice to use the likelihood ratio test and related F-test in astrophysics to compute, e.g., the significance of a weak emission line, those tests are often lacking certain basic conditions and cannot be used for such purposes \citep{Protassov+02, Orlandini12}.
Instead, Bayesian posterior predictive p-values is a robust method \citep{vanDyk2001, Protassov+02} that has recently been favored thanks also to the fact that the computational power it requires is now more largely accessible.
This method uses Monte Carlo simulations of the best-fit model with its parameters uncertainties, to build the reference distribution of the test statistic of choice, and then use that distribution to compute p-values. 
This is often performed in literature through the \texttt{simftest} (to estimate the F-test probability of an additional model component) or the \texttt{fakeit/multifake} (to produce simulated spectra based on real data) tools available in XSPEC.
The procedure adopted by this type of tests and implemented, for example, in \texttt{simftest}, can be summarized as follows:
\begin{enumerate}
    \item fit the observed spectrum, with and without the extra component under investigation, using a test statistic (for example, $\chi^2$); $\Delta \chi^2_{\rm obs}$ from the best fits of the observed spectrum is calculated;
    \item the model without the extra component is considered to be the null hypothesis: $N$ fake spectra based on it (and assuming the same signal-to-noise of the real one) are simulated and then fitted twice: with the same model, with and without the extra component. $\Delta \chi^2$ is calculated for each simulated spectrum;
    \item the obtained $\Delta \chi^2$ can be displayed as a histogram and compared with $\Delta \chi^2_{\rm obs}$ to calculate the probability that the extra component arises by random fluctuations.
\end{enumerate}

Other commonly employed criteria used for models comparison are the Akaike Information Criterion (AIC) \citep{Akaike1974} and the Bayesian Inference Criterion (BIC) \citep{Schwarz1978, Kass1995}, typically employed when the user wishes to penalize extra free parameters (degrees of freedom) in order to avoid overfitting.
See also \citet{Buchner+Boorman2022} for more details on Bayesian methods to test the null hypothesis.

\begin{myblock}{Example of the usage of \texttt{simftest}}
This example is based on \emph{NuSTAR} data of the X-ray binary IGR~J17407$-$2808 (obsid:30801013002), during a low luminosity state \citep[see ][ for more information]{Ducci23}.
Sometimes, to investigate possible relatively narrow features (such as emission lines) in the spectrum, it is advisable to increase the number of bins. If the total number of counts available is too low to use the $\chi^2$ as fit statistic (at least $\sim 25-30$ counts per bin), it is necessary to use another statistic.
\begin{center}
\includegraphics[width=\textwidth]{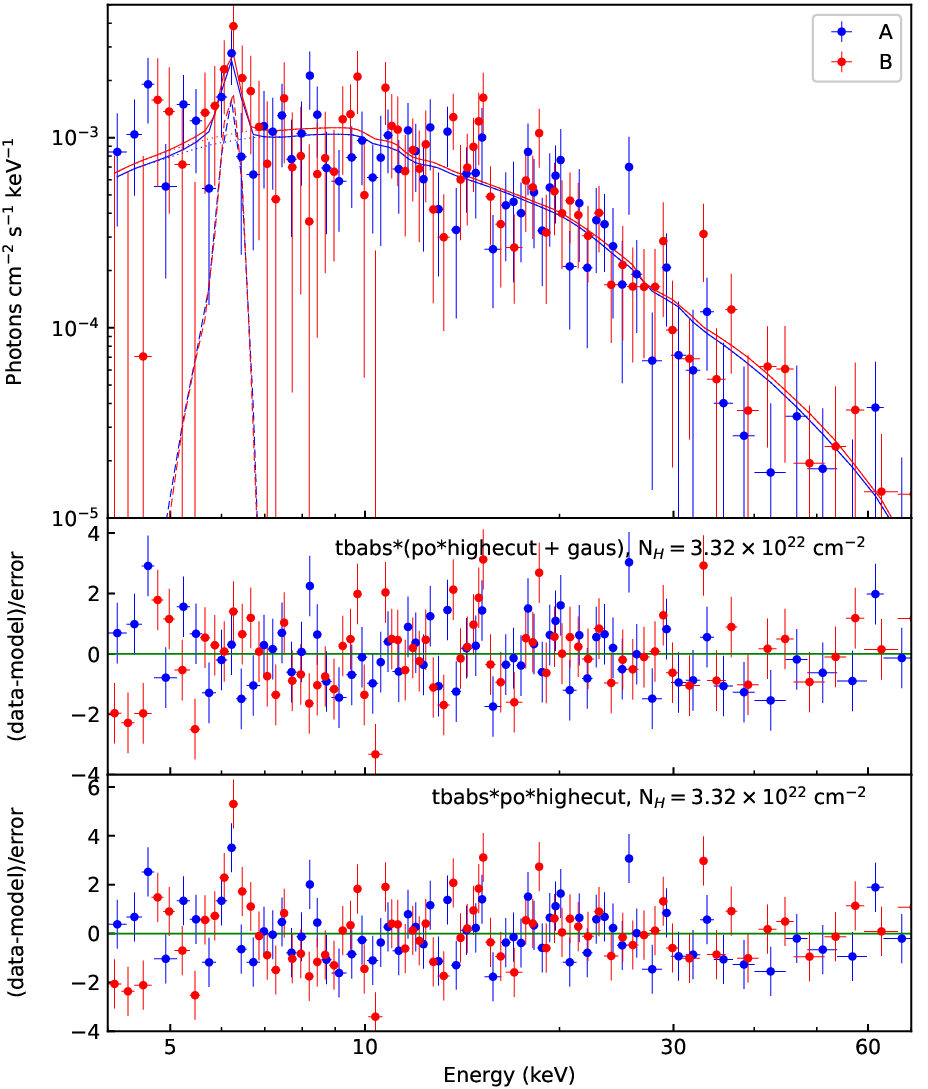}
\captionof{figure}{\emph{NuSTAR} spectra of IGR~J17407$-$2808. \emph{Top panel:} spectra fitted with an absorbed power law with a high-energy cutoff and a Gaussian. \emph{Middle panel:} residuals panel relative to the top panel. \emph{Bottom panel:} residuals obtained by fitting the spectra without the Gaussian component. \label{fig:spec_example_for_simftest}}
\end{center}
In this example, \emph{NuSTAR} module A (black points in Fig. \ref{fig:spec_example_for_simftest}) and B (red points) spectra are grouped to have at least one count per bin, and W fit statistic is adopted. The following commands are used within XSPEC:
\begin{verbatim}
statistic cstat
abund wilm
setplot en
model  constant*TBabs*(powerlaw*highecut)
...
plot lda del
setplot rebin 3 5
plot lda del
setplot add
pl
\end{verbatim}
Here we recall some of the aspects covered in various parts of this chapter, which are relevant for this example. 
In the series of instructions to give to XSPEC above, the abundances were changed to the abundances for the interstellar medium ``wilm'' (the default abundances in XSPEC, ``angr'', would generally overestimate the photoionization cross-section of the interstellar medium; \citealt{Wilms00}).
The goal is to fit two datasets simultaneously, therefore, a cross-calibration normalization constant between NuSTAR/FPMA and FMPB has to be used. The first constant value (linked to data set ``A'') is fixed to 1, the second (linked to data set ``B'') is free to vary (in this case it is expected to be close to 1 because the X-ray source was observed simultaneously by both modules).
Owing to the limited energy coverage at low energies of \emph{NuSTAR}, high degeneracy is expected between the slope of the power law and column density $N_{\rm H}$. The latter has therefore been fixed.
When dealing with low-count spectra having bins grouped to have at least one count per bin (like the one in this example), it might be difficult to visually inspect the overall quality of the fitting results (for example, to look for ``waves'' or evidence of the presence of narrow features in the residuals). In these cases, for a better visual inspection, it is advisable to use the XSPEC command \texttt{setplot rebin}, whose purpose is to change \emph{only} the visualization of the data in the plot (it does not change the real binning of the spectrum: the number of degrees of freedom remain the same; see XSPEC manual for more details).
Returning to the example discussed here, after fitting the data, it is possible to note some residuals at $\sim 6.4$~keV (bottom panel in Fig. \ref{fig:spec_example_for_simftest}), which can be modelled with a Gaussian:
\begin{verbatim}
model  constant*TBabs*(gaussian + powerlaw*highecut)
\end{verbatim}
To calculate the probability that the extra component arises by random fluctuations, \texttt{simftest} can be used (see Sect. \ref{sect: testing model components}):
\begin{verbatim}
simftest 3 10000 simftest_output
\end{verbatim}
In the command above the extra component is the number 3 given as input to XSPEC, that is ``gaus'', and $10^4$ simulations will be performed and the output written in the file \texttt{simftest\_output}. It is possible to plot the results of \texttt{simftest} as an histogram and compared with the $\Delta W$ obtained from real data (Fig. \ref{fig:hist_example_for_simftest}).
\begin{center}
\includegraphics[width=8cm]{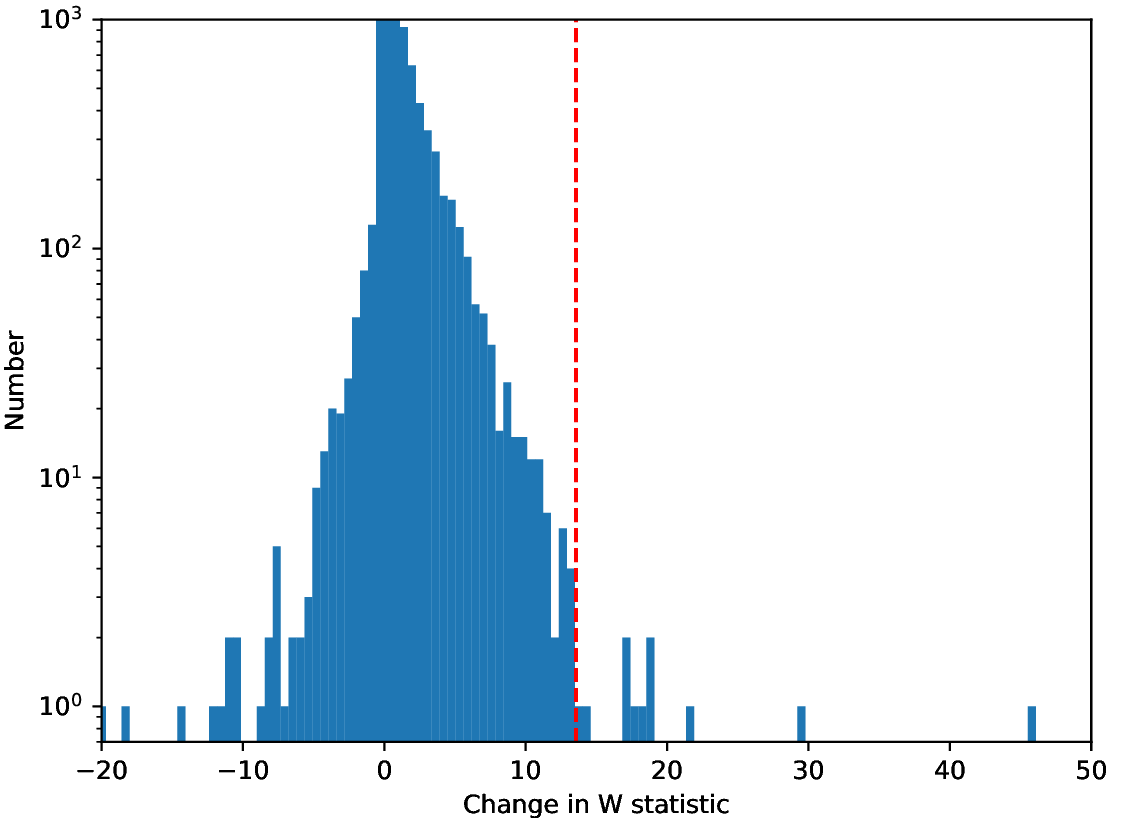}
\captionof{figure}{Results of the simulations performed using \texttt{simftest} for testing the line significance. The histogram shows the $\Delta W$ values obtained in the simulations, compared to the $\Delta W$ from the real data (red vertical line). \label{fig:hist_example_for_simftest}}
\end{center}
The probability that the data are consistent with the spectral model without the Gaussian component is $\sim0.13$\%, which means that the hypothesis of the absence of the emission line cannot be rejected at a $4\sigma$ conﬁdence level. The files to reproduce this example can be retrieved from here: \href{https://github.com/LorenzoDucci/examples_chapter_spectral_analysis}{https://github.com/LorenzoDucci/examples\_chapter\_spectral\_analysis}.
\end{myblock}

\subsection{Parameters correlations and confidence levels}
\label{sect. Parameters correlations and confidence levels}

Another common pitfall to be aware of is the possibility of artificially (i.e., model-driven) correlated parameters.
A typical example of artificial correlation between parameters is that between column density and power law index.
To check if a given set of parameters shows artificial correlation there are several methods.
In this regard, the Fisher matrix is an important indicator of the parameters correlation.
It provides the second derivative of the fit statistic with respect to the parameter at the best-fit solution.
The second derivative value shows how fast the fit statistic increases when it moves away from the best-fit minimum.
For a good parameter estimation, the value of the second derivative is large, meaning that the fit statistic increases rapidly (and the best-fit parameter falls in a true minimum of the fit statistic solution).
The inverse of the Fisher matrix is called the covariance matrix, and it is prompted out by XSPEC at the end of each fit.

Another commonly employed method to assess artificial parameters correlations is to calculate the statistic difference (e.g., $\Delta\chi^2$ or $\Delta$C) from the best-fit for two parameters of interest at the same time.
This is based on the same method employed to calculate parameters errors by varying the statistic of choice S:
\begin{equation}
    S = S_{Best-fit}+\Delta S
\end{equation}
until the $\Delta S$ reaches a critical value correspondent to a certain confidence region.
For two parameters of interests, this method returns iso-statistic surfaces at a given confidence level selected by the $\Delta\chi^2$ or $\Delta$C value for two parameters of interest (which are kept frozen through the fit). 
These surfaces can then be plot as confidence contours.

\begin{table}[h!]
\caption{\label{tab:chi2} Critical $\Delta\chi^2$ corresponding to confidence levels for a given number of parameters of interest \citep{Avni1976}.}
\centering
\begin{tabular}{cccc} 
\hline
Confidence & \multicolumn{3}{c}{\# of Parameters} \\
            & 1 & 2 & 3 \\
\hline
68\% & 1.00 & 2.30 & 3.50 \\
90\% & 2.71 & 4.61 & 6.25 \\ 
99\% & 6.63 & 9.21 & 11.30 \\ 
\hline
\end{tabular}
\end{table}

Typically, the $\chi^2$ fit statistic is employed for this procedure, given that its critical values are known and tabulated as a function of the degrees of freedom (see, e.g., Table~\ref{tab:chi2}).
However, in the higher counts limit ($N>10$), the $\Delta C$ is distributed as $\chi^2$ \citep{Cash79, Wilks38}, and the same critical values can be used to compute confidence levels for the C statistic. 
On the other hand, in the low counts limit, the difference between the two distributions is of the order of $\mathcal{O}(\frac{1}{\sqrt{N}})$ and the above-mentioned approximation is not valid anymore.

Ideally, contour plots would show a circular shape, meaning that no artificial correlation is present between the two parameters. However, some degree of correlation is unavoidable in certain contexts, and the contours show such dependence by inclined ellipsoidal shapes.
When calculating the change in fit statistic, it is also important to adopt the proper stepping size. In fact, if the stepping size is too coarse, or if the initial values of the model parameters are distant from their true best-fit solutions, the fitting procedure might fall in a local minimum of the fit statistic, instead of the true global minimum. To avoid such a pitfall, the stepping size must be chosen in order to finely sample the parameter values around their true best-fit solution. This can be done, e.g., in XSPEC through the \texttt{steppar} tool, which allows to fine-tune the stepping size in a user-defined range of values for one or two parameters of interest.

\begin{figure}
\begin{center}
\includegraphics[width=\textwidth]{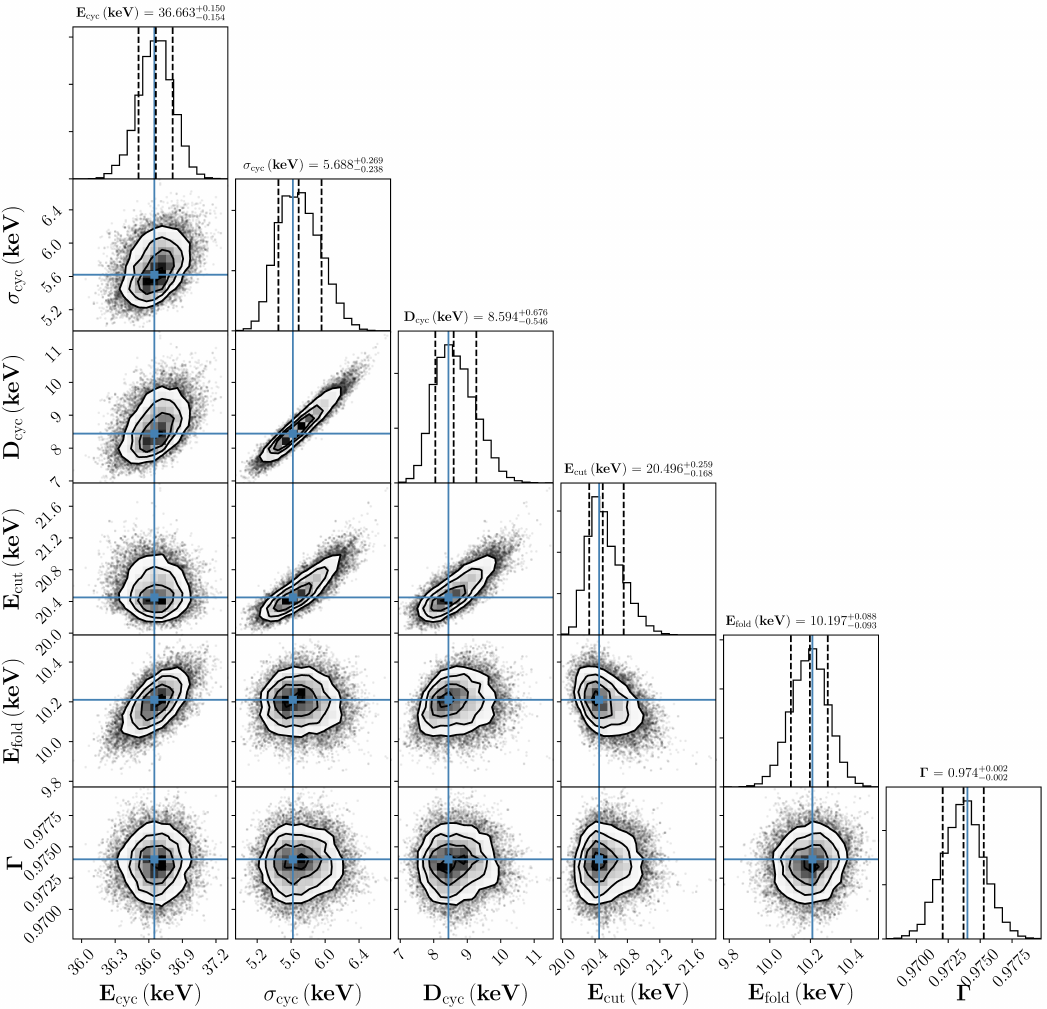}
\end{center}
\caption{Corner plot showing distribution of some of the spectral parameters from the spectral model used to describe the X-ray emission of the low-mass X-ray binary Her~X$-$1 observed by \nustar. The two-dimensional distributions are obtained from MCMC simulations and the histograms on top are the distributions for each parameter. Contours correspond to 68\%, 90\%, and 99\% uncertainty. Blue lines indicate the input parameters used for the simulations (more information in \citet{Staubert20}. Credit: Staubert et al. A\&A 642, A196, 2020, reproduced with permission \copyright ESO.}
\label{fig:mcmc_example}
\end{figure}

An alternative approach to estimate confidence levels relies on Bayes\-ian statistics and is based on Monte Carlo simulations. Although computationally expensive, this approach does not share the usual flaws linked to frequentist statistics, e.g., data rebinning, the assumption that parameters are normally distributed and independent, the necessity of a goodness-of-fit estimator that is model-independent and whose critical probability values are tabulated.
Furthermore, this approach exploits priors to guide the parameter space exploration.
Markov Chain Monte Carlo (MCMC) is a commonly adopted method to determine both the best-fit parameters value and their confidence levels. 
Through MCMC a chain of parameter values is produced, starting from the initial best-fit values (i.e., the proposal distribution). From these, either through the assumption of a given distribution  or by generating multiple sets of walkers \citep{vanDyk2001,GW2010}, a candidate value is first sampled and then tested for acceptance or rejection according to certain conditions, such as the detailed balance condition \citep{Hogg2018}. Once accepted, the sampled value becomes the new starting point from which drawing the next candidate value, and so on, until convergence is reached and a stationary distribution is obtained.
The stationary probability distribution of the spectral parameters can then be used to estimate confidence intervals of parameter uncertainties (see, e.g., Fig.~\ref{fig:mcmc_example}).
Additional details on this approach are given in \citet{Buchner+Boorman2022}.

\subsection{Additional technical recommendations for spectral analysis}

Besides the basic statistics knowledge, additional technical measures need to be implemented for a meaningful fitting procedure.
One of the most important strategies is to start fitting from a suitable initial guess of the parameters values. This is because, in general, the more complicated the model, the more correlated its parameters, and the best-fit solution can fall in a local minimum of the fit statistic, instead of the true global minimum corresponding to the true best-fit solution. Even in the case of simple models, starting the fit from a meaningful value of the parameters will make the computation of the best-fit parameters faster. Overall, it is always recommended to check for global minima and correlation between parameters by using contour plots (see Sect. \ref{sect. Parameters correlations and confidence levels}).

Another common pitfall of spectral analysis regards the interpretation of the final unfolded spectrum.
Inverting Eq.~\ref{eq. CI} would apparently return the unfolded, i.e., incident source spectrum.
However, as outlined in Sect.~\ref{sec:models}, this procedure is unstable and produces uncertain results.
Therefore, unfolded spectra obtained e.g., in XSPEC, are always model-dependent, i.e., they are obtained by substituting the source spectrum $S(E)$ with a model spectrum $M(E)$ in equation Eq.~\ref{eq. CI}.
This is usually done in order to represent the spectrum in a way that, for plotting purposes, is free from response effects.
However, unfolded spectra plots can be deceiving by showing discrete features which are actually a product of the detector response.
For example, the unfolded spectrum can show an emission line where the spectral response presents an absorption edge.
In ISIS, on the other hand, unfolded spectra are defined in a model-independent way \citep{Nowak05}, thus mitigating the influence of the chosen model on the true data points.

In X-ray spectral analysis, it often happens that several data sets need to be fit simultaneously. This may occur, e.g., if a given target has been observed at the same time with different instruments.
In such a case, a simultaneous fit of the spectra obtained with different instruments would improve the statistic and allow overlap among different energy bands.
The standard approach in this case is to tie the model parameters of each data set to their equivalent in the reference data set and add a cross-calibration constant that allows for different normalization values among the various instruments.
If the target is observed at the same time by different instruments, the cross-normalization values among different spectra usually agree within a few percent.

\begin{myblock}{Example of simulations in XSPEC}\label{fig:xcm}

This example illustrates how to use simulations in XSPEC for proposal assessments. In this example, the usage of \texttt{xcm} command text files, the \texttt{fakeit} and \texttt{ftgrouppha}  tools, plus the general understanding of canned response files will be described step by step.
Specifically, this example will deal with \textit{NICER} data.
First, we will assume an approximate knowledge of the spectral model we expect from our source. A typical spectrum is an absorbed power-law with a Gaussian emission line to represent the Iron K$\alpha$ line around 6.4 keV.
We define our model in XSPEC and save it as an \texttt{xcm} file, which will look like this:\\

---------------------------------------------------------------------------\\
method leven 10 0.01\\
abund wilm\\
xsect vern\\
cosmo 70 0 0.73\\
xset delta 0.01\\
systematic 0\\

model  TBabs(cutoffpl + gaussian)\\

        2.0      0.001          0          0     100000      1e+06\\
       0.8       0.01         -3         -2          9         10\\
        15.0       0.01       0.01          1        500        500\\
      0.1       0.01          0          0      1e+20      1e+24\\
        6.4       0.05          0          0      1e+06      1e+06\\
      0.05       0.05          0          0         10         20\\
     0.001       0.01          0          0      1e+20      1e+24\\
     
bayes off\\
%\#\#\#\#\#\#\#\#\#\#\#\#\#\#\#\#\#\#\#\#\#\#\#\#\#\#\#\#\#\#\#\#\#\#\#\#\\
---------------------------------------------------------------------------\\

Once the model is defined, we can simulate a spectrum based on the canned background and response files proper of the instrument we aim to use. 
The command line to do this for \textit{NICER} will look like this:\\

\texttt{fakeit nixtiback20190807.pi \&\\ nixtiref20170601v003.rmf \& \\ nixtiaveonaxis20170601v005.arf \& \\ y \& \& nicer.fak \& 2000}\\

This will produce a simulated \texttt{nicer.fak} spectrum with a 2 ks exposure that still needs to be rebinned before fitting:\\

\texttt{ftgrouppha nicer.fak nicer\_fake\_grp25.pha grouptype=optmin groupscale=25 respfile=nixtiref20170601v003.rmf}\\

At this point we are able to fit our fake spectrum. The model used to seed the simulation is obviously an excellent starting point. 
Moreover, we notice that the background dominates the spectrum for energies below 0.8 keV, so we will ignore the correspondent channels, and those above 10 keV, where \textit{NICER} calibration is not optimal.
Therefore, we can use a modified version of the \texttt{xcm} command file reported above where, besides the model, we can also indicate the spectral data we want to load plus a few other environmental details:\\

---------------------------------------------------------------------------\\
statistic pgstat\\
data 1:1 nicer\_fake\_grp25.pha\\
resp 1:1 nixtiref20170601v003.rmf\\
arf 1:1 nixtiaveonaxis20170601v005.arf\\
backgrnd 1 nixtiback20190807.pi\\
ignore 1:1-11,140-278\\

method leven 10 0.01\\
abund wilm\\
xsect vern\\
cosmo 70 0 0.73\\
xset delta 0.01\\
systematic 0\\

model  TBabs(cutoffpl + gaussian)\\

        1.98588      0.001          0          0     100000      1e+06\\
       0.790526       0.01         -3         -2          9         10\\
         14.652       0.01       0.01          1        500        500\\
      0.0993069       0.01          0          0      1e+20      1e+24\\
        6.36818       0.05          0          0      1e+06      1e+06\\
      0.0620406       0.05          0          0         10         20\\
    0.000824826       0.01          0          0      1e+20      1e+24\\
    
bayes off\\
setplot energy\\
query yes\\
---------------------------------------------------------------------------\\

This will return a $\chi^2$ test statistic of 120 with 121 degrees of freedom and a null hypothesis probability of $\sim52\%$.
Furthermore, we can plot our simulated data and spectral model in order to have a visual representation of our expectations:

\begin{center}
\includegraphics[width=10cm]{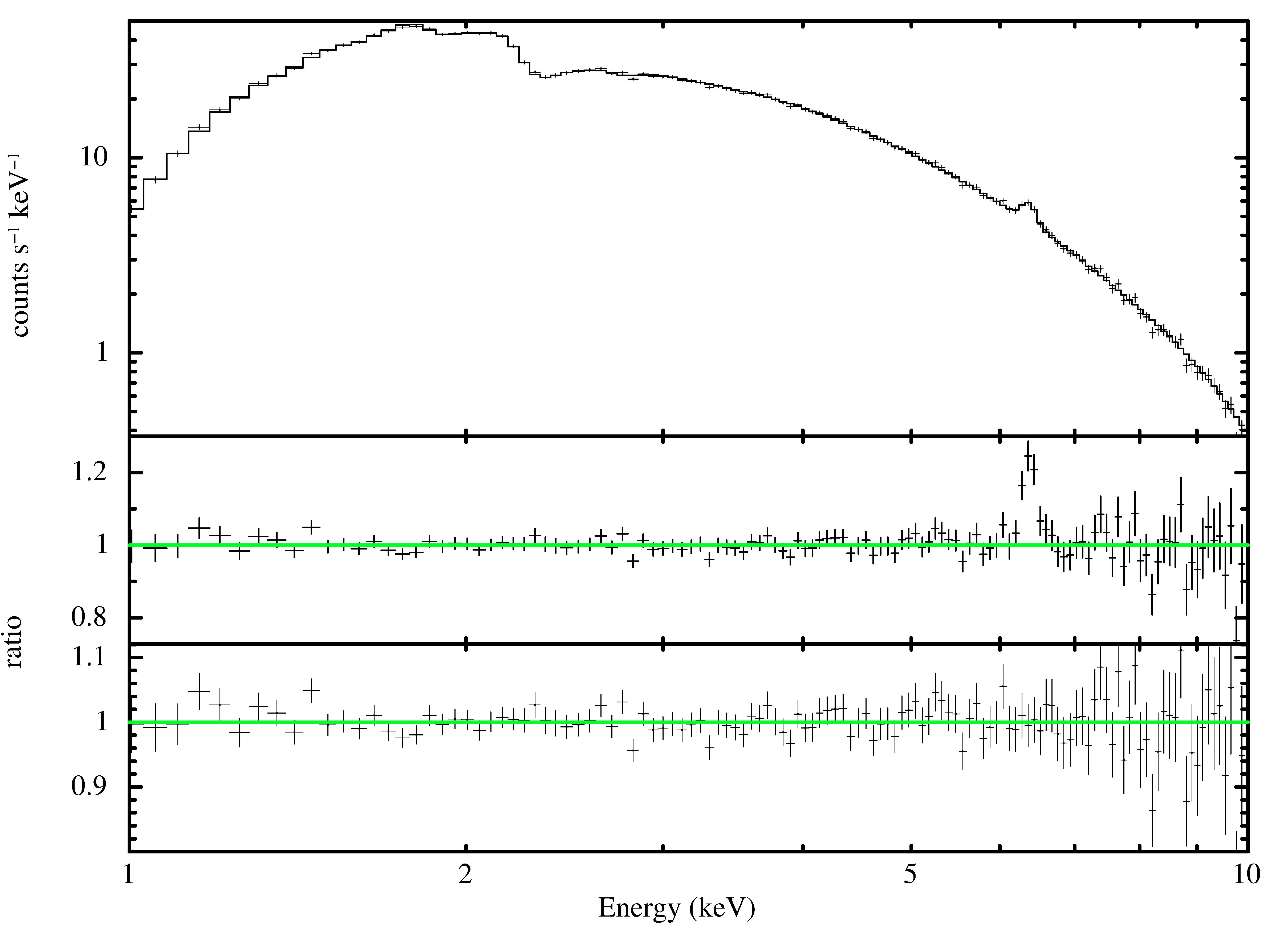}
%\caption{
\captionof{figure}{Example of simulated spectra for NICER with 2 ks exposure. Canned background spectrum and response files have been used to perform the simulation, together with an absorbed cutoff power-law spectrum (see text). \textit{Top panel:} data and folded best-fit model. \textit{Middle panel:} residuals from the best-fit model without an Iron K$\alpha$ line. \textit{Bottom panel:} Residuals from the best-fit model including an Iron K$\alpha$ line around 6.4 keV.%}
\label{fig:simulated_nicer}}
\end{center}

\normalsize The residual panels in Fig. \ref{fig:simulated_nicer} clearly show how different the models with and without the Gaussian line component are. 
In the case without the inclusion of the Gaussian line, residuals reach up to $\sim25\%$ of the model, thus indicating that the line is highly significant.
In case of, e.g., the preparation of an observational proposal, to quantify the significance of the line a new set of simulations similar to those plotted in Fig. \ref{fig:hist_example_for_simftest} can be run in order to assess if the assumed exposure time (2~ks in our case) is sufficient to detect the feature within the user-defined detection threshold (e.g., $3\sigma$). 
If not, the exposure time can be increased in steps until the desired detection threshold is reached.
On the opposite, if the significance of the line appears unnecessarily high, the exposure time can be lowered until the user requirements are met but without jeopardizing the allocation of the proposal due to unjustified requested time.

The files to reproduce this example can be retrieved from here:

\noindent \href{https://github.com/LorenzoDucci/examples_chapter_spectral_analysis}{https://github.com/LorenzoDucci/examples\_chapter\_spectral\_analysis}.\\
\end{myblock}

\section{Performance estimates for proposals and surveys}\label{sec:performance}

One of the most important aspects of an observing proposal is an appropriate study of the technical feasibility of the observation. The first question to be answered is whether the target can be detected by an X-ray telescope within a given exposure time. The most used tool for this purpose is the Portable, Interactive, Multi-Mission Simulator (PIMMS, and its web version webPIMMS\footnote{\url{https://heasarc.gsfc.nasa.gov/cgi-bin/Tools/w3pimms/w3pimms.pl}}, \citealt{Mukai93}). The software is able to convert rates and fluxes detected by several X-ray missions, given simple spectral models, which take into account interstellar absorption and continuum components such as power-law, blackbody, bremsstrahlung, Gaussian, and a model for collisionally-ionized diffuse gas. 
For some missions (e.g., \emph{NICER} and \emph{NuSTAR}), PIMMS and webPIMMS provide the exposure time required to achieve a $5\sigma$ detection\footnote{The output energy range must be set as 'default'.}. For other missions (e.g., \emph{Chandra} and \xmm), these tools also provide useful information for preparing the feasibility of an observing proposal, such as a pile-up estimate.
In this regard, spectral simulations are of great aid. For example, when a new instrument for spectroscopy is designed, its response matrix can be used to simulate spectra to demonstrate
the performance of a new instrument. 
Another reason to simulate spectra is to demonstrate the feasibility
of a proposed observation, through the estimate of the exposure time needed to reach an observational goal, and to determine the constraints on the spectral parameters of the proposed observation.
A well-known online tool designed to this aim is WebSpec\footnote{\url{https://heasarc.gsfc.nasa.gov/webspec/webspec.html}}. 
With WebSpec the user can simulate a certain spectral model as obtained through a given instrument, thus assessing the scientific return of the planned observation.
Moreover, some of the available software for spectral modeling (e.g., XSPEC) offer the possibility to simulate X-ray spectra, assuming as input an exposure time, a spectral model, the instrument response files (ARF and RMF) and possibly a background file. Once the simulated data are created, they can be analyzed by the same software as real data.
An example is given on page~\pageref{fig:xcm}, where a simulated NICER spectrum is obtained with the \texttt{fakeit} tool in XSPEC, assuming a 2 ks exposure and an absorbed cutoff power-law model with a Gaussian emission line at 6.4 keV. 
Part of the correspondent header of the simulated spectrum is shown in Table~\ref{table:header}.
The simulation can be used to test, e.g., if a highly-significant Iron K$\alpha$ emission line at 6.4 keV is detected within a given exposure time.
For a more statistically robust demonstration, the \texttt{multifake} tool can be used to test if a lower-significance model component can also be detected and to derive confidence levels for the component best-fit parameters.

Likewise, spectral simulations are also used to explore surveys capabilities. In fact, spectral models are fundamental in exploiting X-ray surveys and the instrument sensitivity either for quick scans or for deep fields. This is due to the fact that a certain source's detection limit within a given exposure time and in a given energy band depends on the source and background spectral shape. Thus, spectral simulations help driving surveys achievements and maximize the impact of surveys.

\acknowledgement{LD acknowledges support from the \textsl{Bundesministerium f\"{u}r Wirtschaft und Energie} through the \textsl{Deutsches Zentrum f\"{u}r Luft- und Raumfahrt e.V. (DLR)} under the grant number FKZ 50 QR 2102.}

\bibliographystyle{aasjournal}
\bibliography{main}

\newpage

\end{document}